\documentclass[aps,prd,twocolumn,showpacs,superscriptaddress,
               amsmath,amssymb]{revtex4-1}

\usepackage{dcolumn}
\usepackage{bm}
\usepackage{graphicx}
\usepackage{color}
\usepackage{dsfont}
\usepackage{bbm}
\usepackage{amsmath}
\usepackage{amssymb}
\usepackage{float}
\usepackage{psfrag}
\usepackage{mathdots}

\def\urltilda{\kern -.15em\lower .7ex\hbox{\~{}}\kern .04em}

\graphicspath{{figures/}}

\begin{document}

\title{Photon Torpedoes and Rytov Pinwheels:
       Integral-Equation Modeling of Non-Equilibrium
       Fluctuation-Induced Forces and Torques
       on Nanoparticles}

\author{M. T. H. Reid}
\affiliation{Department of Mathematics, Massachusetts Institute of Technology, Cambridge, MA 02139}
\author{O. D. Miller}
\affiliation{Department of Applied Physics and Energy Sciences Institute, Yale University, New Haven, CT 06511}
\author{A. G. Polimeridis}
\affiliation{Skolkovo Institute of Science and Technology, Moscow, Russia}
\author{A. W. Rodriguez}
\affiliation{Department of Electrical Engineering, Princeton University, Princeton, NJ}
\author{E. M. Tomlinson}
\affiliation{Department of Physics, Massachusetts Institute of Technology, Cambridge, MA 02139}
\author{S. G. Johnson}
\affiliation{Department of Mathematics, Massachusetts Institute of Technology, Cambridge, MA 02139}

\date{\today}


\newcommand\wt{\widetilde}
\newcommand\vbChi{\boldsymbol{\chi}}
\newcommand\vbphi{\boldsymbol{\phi}}
\newcommand\vbPhi{\boldsymbol{\Phi}}
\newcommand\vbDelta{\boldsymbol{\Delta}}
\newcommand\vbLambda{\boldsymbol{\Lambda}}
\newcommand\vbGamma{\boldsymbol{\Gamma}}
\newcommand\vbOmega{\boldsymbol{\omega}}
\newcommand\vbrho{\boldsymbol{\rho}}
\newcommand\vbchi{\boldsymbol{\chi}}
\newcommand\vbEps{\boldsymbol{\epsilon}}
\newcommand\vbMu{\boldsymbol{\mu}}

\newcommand{\red}[1]{\textcolor{black}{#1}}

\newcommand{\ket}[1]{\left|#1\right\rangle}
\newcommand{\Ket}[1]{\Big|#1\Big\rangle}
\newcommand{\KET}[1]{\bigg|#1\bigg\rangle}
\newcommand{\bra}[1]{\right<#1\left|}
\newcommand{\Bra}[1]{\Big<#1\Big|}
\newcommand{\BRA}[1]{\bigg<#1\bigg|}
\newcommand{\ExpVal}[1]{\left\langle#1\right\rangle}
\newcommand{\ExpValb}[1]{\big\langle#1\big\rangle}
\newcommand{\ExpValB}[1]{\Big\langle#1\Big\rangle}
\newcommand{\inp}[2]{\left\langle#1\right|\left.#2\right\rangle}
\newcommand{\Inp}[2]{\big\langle#1\big|\big.#2\big\rangle}
\newcommand{\INP}[2]{\Big\langle#1\Big|\Big.#2\Big\rangle}
\newcommand{\exptwo}[3]{\left\langle#1\right|#2\left|#3\right\rangle} 
\newcommand{\Exptwo}[3]{\big\langle#1\big|#2\big|#3\big\rangle}
\newcommand{\EXPTWO}[3]{\Big\langle#1\Big|#2\Big|#3\Big\rangle}
\newcommand\unitmatrix{\mathds{1}}
\newcommand\Tr{\hbox{Tr }}
\newcommand\sups[1]{^{\hbox{\scriptsize{#1}}}}
\newcommand\supt[1]{^{\hbox{\tiny{#1}}}}
\newcommand\subs[1]{_{\hbox{\scriptsize{#1}}}}
\newcommand\subt[1]{_{\hbox{\tiny{#1}}}}
\newcommand{\nn}{\nonumber \\}
\newcommand{\vb}[1]{\mathbf{#1}}
\newcommand{\mb}[1]{\mathbb{#1}}
\newcommand{\numeq}[2]{\begin{equation} #2 \label{#1} \end{equation}}
\newcommand{\pard}[2]{\frac{\partial #1}{\partial #2}}
\newcommand{\pardn}[3]{\frac{\partial^{#1} #2}{\partial #3^{#1}}}
\newcommand{\pf}[2]{\left(\frac{#1}{#2}\right)}
\newcommand{\vbhat}[1]{\vb{\hat #1}}
\newcommand{\vbhatt}[1]{\boldsymbol{\widehat #1}}
\newcommand{\mc}[1]{\mathcal{#1}}
\newcommand{\bmc}[1]{\boldsymbol{\mathcal{#1}}}
\newcommand{\citeasnoun}[1]{Ref.~\citenum{#1}}
\newcommand{\citeasnouns}[2]{Refs.~\citenum{#1},~\citenum{#2}}
\newcommand{\wh}{\widehat}
\newcommand{\jBar}{\overline{j}}
\newcommand{\hBar}{\overline{h}}
\newcommand{\jSlash}{\backslash\hspace{-0.07in}j}
\newcommand{\hSlash}{\backslash\hspace{-0.07in}h}

\begin{abstract}

We present new theoretical tools, \red{based on fluctuational
electrodynamics and the integral-equation approach
to computational electromagnetism,}
for numerical modeling of forces and torques
on bodies of complex shapes and materials \red{due to emission
of thermal radiation out of thermal equilibrium.
This extends our recently-developed fluctuating-surface-current (FSC) and
fluctuating-volume-current (FVC) techniques for radiative heat transfer
to the computation of non-equilibrium fluctuation-induced
forces and torques; as we show, the extension is non-trivial
due to the significantly greater computational cost of
modeling radiative momentum transfer, including a new
singularity not present in the energy-transfer case
that must be carefully neutralized to yield a tractable
solver. We introduce a new analytical cancellation technique
that addresses these challenges and allows, for the first
time, accurate and efficient prediction of non-equilibrium
forces and torques on bodies of essentially arbitrary
shapes---including asymmetric and chiral particles---and
complex material properties, including continuously-varying
and anisotropic dielectrics.}
We validate our approach by showing that it reproduces known results, 
then present new numerical predictions of non-equilibrium
self-propulsion, self-rotation, and momentum-transfer phenomena
in complex geometries that would be difficult or impossible to study
with existing methods.
Our findings indicate that the fluctuation-induced dynamics of
micron-size room-temperature bodies in cold environments
involvee microscopic length scales but macroscopic time scales, 
with typical linear and angular velocities on the order of 
microns/second and radians/second; moreover,
for particles of fixed shape \red{we find} an
optimum particle size at which linear or angular
acceleration is maximized.
For a micron-scale gear driven by absorption of angular momentum from
the thermal radiation of a nearby chiral emitter, we find a strong 
and non-monotonic dependence 
of the magnitude and even the \textit{sign} of the induced torque
on the temperature of the emitter,
suggesting applications to precision nanoscale thermometry.
\end{abstract}

\maketitle

\section{Introduction}
\label{IntroductionSection}
This paper presents new theoretical tools
for solving a century-old problem: predicting
fluctuation-induced forces and torques on material
bodies originating from asymmetric emission of thermal
radiation, including thermal self-propulsion/self-rotation
(TSP/TSR) of isolated bodies and non-equilibrium (NEQ)
Casimir forces/torques between bodies. Although these
effects have captured the imagination of scientists and
engineers for over a century~\cite{Crookes1874, Rubincam2007,
Beekman2006} and have been observed in experiments spanning a 
vast range of length scales---from laboratory measurements of 
nanofabricated devices~\cite{Liu2010} to astronomical 
observations of natural~\cite{Kaasalainen2007} and 
man-made~\cite{Turyshev2012} celestial bodies---the science of 
\textit{theoretical} modeling of NEQ fluctuation-induced
forces and torques has remained in its infancy due to a
host of forbidding technical challenges.
This may seem surprising in view of
rapid recent progress in theoretical methods
for the closely related problems of Casimir forces/torques
(fluctuation-induced momentum transfer in thermal
equilibrium)~\cite{Reynaud2006, Wagner2008, Scheel2009, Rahi2009, Johnson2011}
and near-field radiative heat exchange
(fluctuation-induced energy transfer out of thermal
equilibrium)~\cite{Narayanaswamy2008,Messina2011,Otey2011,Krueger2012,Lussange2012, Rodriguez2012, Polimeridis2015B}; 
\begin{figure}[!h]
\begin{center}
\resizebox{0.5\textwidth}{!}{\includegraphics{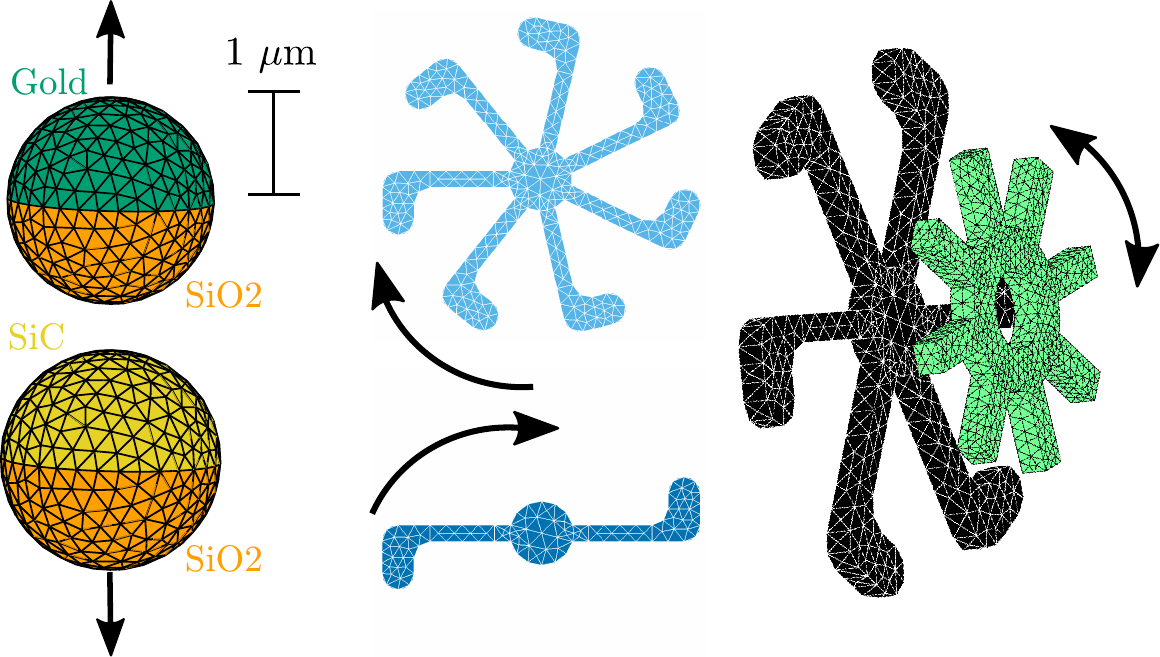}}
\caption{Synopsys of main results. \textit{Photon torpedoes:}
\red{
Warm ``Janus'' particles~\cite{Walther2013, Ilic2017} consisting
of conjoined hemispheres of distinct media self-propel in different 
directions depending on the material combination}
(Section \ref{PhotonTorpedoSection}).
\textit{Rytov pinwheels:} Warm chiral particles in
a cold environment
self-rotate counter-clockwise or clockwise depending on the
number of pinwheel arms (Section \ref{RytovPinwheelSection}).
\textit{Non-contact microgears:}
A warm Rytov pinwheel causes a nearby gear to spin
clockwise or counterclockwise depending on 
temperature and material combination (Section \ref{MicroGearSection}).
All structures have linear dimensions on the order of $1\, \mu$m.
}
\label{DigestFigure}
\end{center}
\vspace{-0.4in}
\end{figure}
although results have recently appeared for NEQ forces
in geometries involving spheres and/or smooth or corrugated 
plates~\cite{Krueger2011, Messina2011, Krueger2012,
Messina2014, Zheng2015, Muller2016}, 
we explain below why the extension to NEQ forces/torques
on \textit{arbitrary} bodies is nontrivial, requiring
new ideas beyond what is required for equilibrium or
heat-transfer problems.
A key challenge is the emergence of new divergences
that must be carefully neutralized to obtain a tractable
numerical method (Section 2).
\red{We introduce a new analytical cancellation that solves 
this problem}
(Section 3) and present efficient, general-purpose
algorithms for accurately predicting TSP/TSR and NEQ
forces/torques on bodies of complex shapes and material
properties; our tools, which are available as free,
open-source software~\cite{scuff-em, buff-em},
extend both the
\textit{fluctuating-surface-current}~\cite{Reid2013B, Rodriguez2013B}
and \textit{fluctuating-volume-current}~\cite{Polimeridis2015B}
approaches to fluctuation-induced interactions, which we
previously used to study EQ Casimir
forces~\cite{Reid2009} and NEQ heat
transfer~\cite{Rodriguez2013A, Polimeridis2015B}.
After validating our methods by reproducing
known results for non-equilibrium Casimir interactions
(Section 4), we use our new tools to
obtain numerical
predictions of novel fluctuation-induced force and torque
phenomena in complicated geometries that would be practically
impossible to treat using existing methods (Section 5).
Among our findings are that micron-scale asymmetric bodies
at room temperature, initially at rest in cold vacuum
environments, accelerate while radiatively cooling, ramping
up over a time on the order of seconds to terminal linear 
or angular velocities on the order of microns/second or
radians/second;
moreover, varying the overall size of fixed-shape particles
reveals well-defined optimal sizes for maximal self-thrust.

Thermal and quantum-mechanical fluctuations in
material bodies give rise to radiated fields which
carry both \textbf{(a)} energy and \textbf{(b)}
momentum to nearby bodies and
the surrounding environment~\cite{Landau1, Landau2}.
Although mathematical modeling of phenomenon \textbf{(a)}
dates back to the 19th-century theory of
black-body radiation~\cite{Landau1}
and that of phenomenon \textbf{(b)} to the mid-20th century
theory of Casimir phenomena~\cite{CasimirPolder1948, Casimir1948},
recent years have witnessed dramatic advances in the power
of theoretical tools to make rigorous predictions
of fluctuation-induced phenomena in geometries
involving bodies of complex shapes and 
materials~\cite{Reynaud2006, Wagner2008, Scheel2009, Rahi2009,
Lambrecht2011,Johnson2011, Narayanaswamy2008,Messina2011,
Otey2011,Krueger2012, Intravaia2012,
Lussange2012,Reid2013B,Rodriguez2012,Polimeridis2015B}.
In many cases, these tools---which include both
scattering-theoretic methods yielding elegant 
analytical formulas for high-symmetry
geometries~\cite{Scheel2009, Rahi2009,Lambrecht2011,Narayanaswamy2008, 
Messina2011,Otey2011,Krueger2012,Lussange2012}
and fully numerical methods capable of studying
arbitrary-shaped bodies~\cite{Reid2009, Pasquali2009,
Johnson2011, Rodriguez2012,
Reid2013B, Polimeridis2015B}---were originally
developed to study equilibrium momentum-transfer
(Casimir) phenomena~\cite{Rahi2009,Lambrecht2011,Reid2013B}
and later extended to handle non-equilibrium
energy transfer~\cite{Krueger2012,Messina2011,
Rodriguez2012,Polimeridis2015B}.

The further extension to handle non-equilibrium
\textit{momentum} transfer would seem a
logical next step, and indeed some
scattering-matrix methods have been
generalized to handle non-equilibrium
forces~\cite{Krueger2011, Messina2011,
Krueger2012, Messina2014, Zheng2015, 
Muller2016},
and self-propulsion~\cite{Muller2016}
in geometries involving spheres
and/or smooth or corrugated plates.
However, such studies are relatively
few in comparison to the dozens of
equilibrium Casimir and heat-transfer
geometries that have been investigated
by scattering-matrix methods~\cite{Wagner2008,
Narayanaswamy2008,
Rahi2009,
Lambrecht2011,
Messina2011,
Otey2011,
Krueger2012,
Intravaia2012,
Lussange2012, 
Messina2015,
Guerout2015};
among the reasons for this may be that
the arena in which scattering-matrix methods
are most powerful---modeling of high-symmetry
bodies---is the maximally uninteresting regime
for studies of self-propulsion and self-rotation,
which rely crucially on \textit{asymmetry}
and vanish for symmetric geometries such
as a homogeneous sphere in 
isolation~\cite{Krueger2011, Muller2016}.
In principle, this should be the natural entry
point for numerical methods~\cite{Johnson2011} 
such as finite-difference~\cite{Pasquali2009}
or integral-equation~\cite{Reid2013B}
approaches, which treat all
bodies---symmetric or otherwise---on an equal
footing and have been used to study
EQ-force and heat-transfer problems in
complex geometries for which scattering-matrix
methods would be unwieldy~\cite{Johnson2011, Reid2013A}.
However, as discussed in detail below
(Section \ref{WhyHardSection}), the extension
of these methods to non-equilibrium momentum-transfer
problems turns out to be highly nontrivial due to
several complications, including 
\textbf{(a)} rapidly oscillatory functions of position
or frequency that pose high costs for numerical integration,
and \textbf{(b)} a new divergence,
stemming ultimately from the fact that electrostatic
fields carry momentum but not energy~\cite{Jackson1999},
that does not affect EQ-force or heat-transfer problems
but must be carefully isolated and cancelled
analytically to avoid burying NEQ-force signals in
numerical noise.

Our resolution of these difficulties is presented
in Section \ref{TheorySection},
where we present fluctuating-surface-current (FSC)
and fluctuating-volume-current (FVC) formulas
for non-equilibrium momentum exchange among
bodies of arbitrary shapes and materials and
with their environment.
The FSC method, which is based on the
surface-integral-equation (SIE) formulation of classical 
electromagnetic scattering~\cite{Harrington1996, Chew2009, Volakis2012},
originated as a technique for modeling equilibrium
Casimir forces between bodies of homogeneous, isotropic 
material properties~\cite{Reid2009, Reid2011A, Reid2013B}
and was subsequently extended to describe non-equilibrium
heat transfer between such
bodies~\cite{Rodriguez2012, Rodriguez2013B}.
Later, the FVC approach to heat transfer~\cite{Polimeridis2015B},
which is based on the volume-integral-equation (VIE) approach
to classical scattering~\cite{Harrington1996, Chew2009, Volakis2012},
was developed as an extension of the FSC paradigm
to inhomogeneous and anisotropic bodies.

Although the FSC and FVC techniques differ in their description
of currents in material bodies and require completely separate
numerical implementations, they share a common conceptual
foundation, which was discussed separately for the two cases
in~\citeasnouns{Polimeridis2015B}{Rodriguez2013B}
for the
specific problem of heat transfer. In
Section \ref{CommonTheorySection} we extend
those discussions to more general NEQ-fluctuation
problems---including force and torque prediction---using
a notation that emphasizes the common structure of the
FSC and FVC formulas for fluctuation-induced powers, forces,
and torques (PFTs).
These formulas involve the traces of products of certain
matrices describing the interactions of currents in
material bodies; Sections \ref{FVCTheorySection} and
\ref{FSCTheorySection} respectively discuss the different 
forms assumed by these matrices in the FVC and FSC cases,
and Section \ref{QPFTSection} discusses
a crucial analytical cancellation required in both cases
to yield reliable numerical results for NEQ forces and 
torques. The upshot is a single set of master matrix-trace
formulas [equations (\ref{MasterFormulas}) in Section
\ref{MasterFormulasSection}] that may be used to
compute fluctuation-induced PFTs in both the FSC and FVC
contexts. 

An advantage of FSC/FVC methods is that they are agnostic
with respect to the choice of basis functions used to describe
currents; as discussed in Section \ref{BasisFunctionSection},
there are multiple choices of basis in which
to evaluate the matrix-trace formulas (\ref{MasterFormulas}).
For numerical studies of complex, asymmetric geometries,
it is convenient to follow common practice in the
computer-aided design community~\cite{Harrington1996, Chew2009}
by using localized basis functions conforming to a
geometry mesh~\cite{RWG1982,SWG1984}; this is the approach
used in all previous FSC and FVC studies~\cite{Reid2009,
Reid2013B, Rodriguez2013B, Polimeridis2015B},
and we adopt it as our strategy in this paper,
using meshed geometries with localized basis
functions both to validate our new tools by comparison
with known results (Section \ref{ValidationSection})
and to explore novel self-propulsion and self-rotation
effects in micron-scale bodies of complex shapes
(Section \ref{ApplicationsSection}).
%

The FSC/FVC approach may be applied to predict fluctuation-induced
momentum-transfer phenomena in geometries consisting of any number 
of bodies.
In Section \ref{ValidationSection} we validate our methods
by showing that they correctly reproduce known results for
non-equilibrium forces between two homogeneous
spheres~\cite{Krueger2012}. 
Thereafter, we focus on the case of individual
bodies, using FSC and FVC techniques to obtain
new numerical predictions\red{---summarized schematically 
in Figure \ref{DigestFigure}---}for
homogeneous and inhomogeneous nanoparticles
exhibiting thermal self-propulsion (``photon torpedoes,''
Section \ref{PhotonTorpedoSection}),
thermal self-rotation (``Rytov pinwheels,''
Section \ref{RytovPinwheelSection}),
and mechanical actuation mediated by
thermal radiation (non-contact thermal microgears,
Section \ref{MicroGearSection}).
We show that the dynamics of these particles involves
microscopic length scales but macroscopic time scales:
a warm micron-scale asymmetric body placed at rest in a cold 
environment ramps up to a terminal linear or
angular velocity on the order of microns/second 
or radians/second (Figures 2-4), and varying the overall
scale of particles with fixed shapes identifies a
optimal size at which thermal self-acceleration
is maximized.

The formulas used in this paper to model 
fluctuation-induced phenomena are implemented 
in {\sc scuff-em}~\cite{scuff-em}
and 
in {\sc buff-em}~\cite{buff-em}, 
free, open-source software implementations
of the FSC and FVC approaches to 
fluctuational electrodynamics.

\section{Why are non-equilibrium forces and torques hard to compute?}
\label{WhyHardSection}

As noted above, recent years have seen the emergence
of sophisticated computational techniques for modeling
fluctuation-induced
\textit{momentum} exchange \textit{in} thermal equilibrium
(Casimir forces/torques)~\cite{Reynaud2006, Wagner2008, Rahi2009, Johnson2011}
and fluctuation-induced \textit{energy} exchange 
\textit{out of} equilibrium (radiative heat 
transfer)~\cite{Narayanaswamy2008,Messina2011,Otey2011,Krueger2012,Lussange2012, Rodriguez2012, Polimeridis2015B},
spurring predictions of novel EQ Casimir phenomena and NEQ 
heat-transfer phenomena in a wide variety of complex 
geometries.
In contrast, studies of NEQ forces/torques
have been fewer, and have been restricted thus far to
relatively high-symmetry geometries---spheres
and/or and/or smooth or corrugated plates~\cite{Krueger2011,
Messina2011, Krueger2012, Messina2014, Zheng2015, Muller2016}---that 
admit geometry-specific scattering-matrix 
treatments~\cite{Narayanaswamy2008,Messina2011,Otey2011,Krueger2012,Lussange2012}.
Why has it been so difficult to extend existing numerical
solvers for fluctuations in general geometries to the
problem of NEQ momentum transfers? The goal of this
section is to offer a qualitative flavor of the answer
to this question, summarizing the computational challenges 
involved with details relegated to later sections.
As we explain, extending existing tools to the problem 
of NEQ forces in arbitrary geometries is \textit{not}
simply a matter of adding an NEQ module to an EQ code,
or tacking a force module onto a heat-transfer code;
instead, new ideas are required to address
unique problems that arise.

At an intuitive level, the greater computational
complexity of NEQ forces vs. powers is suggested already 
by elementary thought experiments. Consider a warm body
at uniform temperature $T$ in a cold environment.
A simple way to model the energy and momentum lost
to thermal radiation is to assume that photons emitted 
from each infinitesimal surface patch $dA$ carry away 
energy $dE=\alpha \,dA$ and momentum 
$d\vb P=\frac{\alpha}{c} \vbhat{n} \,dA$
per unit time~\cite{Jackson1999}; here $\alpha$ is a 
temperature- and material-dependent but position-independent
constant, $c$ is the speed of light, and $\vbhat{n}$
is the outward-pointing normal to the surface patch.
The total rate of energy loss
$P=\alpha \int dA = \alpha A$ then scales with surface
area, a prediction that captures the correct
qualitative behavior~\cite{Landau1}
and is in quantitative
agreement with experimental observations across
length scales from the macroscopic to the
micron-scale~\cite{Rytov1987},
although corrections are required
for micron-scale or smaller bodies~\cite{PolderVanHove1971}.
On the other hand, the total \textit{momentum}
loss per unit time \textit{vanishes} in this 
simple picture, $-\vb F=\frac{\alpha}{c}\int_A \vbhat{n} dA=0$,
whereupon the \textit{entirety} of the force arises
from higher-order corrections; the simple
physical model that captures the essential
features of radiative energy loss is of no 
predictive power for the force. (This is for bodies of 
uniform temperature; in the presence of thermal
gradients the na\"ive picture of normal-directed
momentum emission \textit{can} suffice to yield
accurate force estimations~\cite{Turyshev2012}.)

The distinction between power and force here
is essentially due to the vector nature of
momentum: whereas the \textit{energy} radiation from
different regions of a warm body is a
scalar quantity that \red{typically} adds constructively over
the surface, the linear and angular \textit{momentum}
emissions vary in direction over the surface
and exhibit significant cancellation \red{in many cases},
requiring costly fine-grained sampling to resolve 
numerically~\cite{Jackson1999}. This problem confounds
attempts to compute fluctuation-induced forces on bodies 
by numerically integrating the thermally-averaged
Maxwell stress tensor (MST) over bounding surface
enclosing the body (Section IIIa below); cancellations 
from different surfaces must be carefully resolved,
at high computational cost, \red{even as numerical
quadrature of the Poynting flux involves little or
no cancellation and generally converges rapidly to
yield accurate heat-radiation rates~\cite{Mishchenko2002}.}

In principle, this difficulty could be overcome simply
by sampling the integrand at large numbers of cubature
points on the bounding surface~\cite{Cools2003}, and 
such a brute-force 
approach is probably tractable for integral-equation 
approaches to classical, deterministic problems~\cite{Harrington1996, Chew2009}:
the MST at each cubature point is a quadratic function
of the $\vb E$ and $\vb H$ fields there~\cite{Jackson1999},
and these in turn are weighted superpositions of the 
elemental fields radiated by the current distributions of
individual basis functions, so the cost of numerical 
MST integration with $N\subt{C}$ cubature points in a
geometry with $N\subt{BF}$ basis functions scales
like $O(N\subt{C}N\subt{BF})$, an inexpensive 
post-processing step for integral-equation solvers.
But things are not so easy for fluctuation-induced 
phenomena, where we must evaluate the statistical
\textit{average} of the MST at each point~\cite{Landau2},
requiring
thermally-averaged products of field components; as
discussed below (Section IIIa), the quantities that 
superpose here are not the fields of individual basis 
functions but the \textit{products} of fields of individual 
basis functions, of which---for each cubature point---there
are $\sim N\subt{BF}^2$. Each additional cubature point
thus increases computational cost by an enormous 
amount proportional to $N\subt{BF}^2$, and the task of
resolving delicate MST cancellations by fine-grained
sampling quickly becomes intractable.

Of course, surface integration of the MST is not
the only way to compute momentum transfer to a body;
an equivalent alternative is to consider \textit{volume} 
integration of the local Lorentz forces~\cite{Jackson1999}
on
charges and currents in the body (Section IIIb,c below).
This technique, which is the basis of the scattering-matrix
approach to NEQ forces of~\citeasnoun{Krueger2012},
is closely analogous
to the technique of computing power transfer by
volume-integrating the local Joule heating, as is
done in both scattering-matrix approaches~\cite{Krueger2012}
and numerical FSC and FVC solvers for NEQ heat
transfer~\cite{Rodriguez2012, Polimeridis2015B}. 
However, in attempting to apply
numerical volume-integral methods to calculate
forces and torques in arbitrary-shaped bodies,
one runs into yet another problematic manifestation
of the distinct character of electromagnetic
momentum vs. energy transfer---namely,
that quasistatic fields produce forces but 
not power transfer~\cite{Landau3}. An obvious example is 
furnished by two monochromatic (frequency $\omega$) 
point dipoles, which exert a force on each
other that remains finite as $\omega \to 0$
even as the energy exchange vanishes in
that limit~\cite{Landau3}.
Similarly, a pointlike dipole radiator
at a distance $d$ from the surface of a material
body transfers energy to the body at a rate that
remains finite as $d\to 0$, but exerts a
\textit{force} that \textit{diverges} in this
limit due to the contribution of quasistatic 
fields~\cite{Jackson1999}.

It is this latter phenomenon that is troublesome
in numerical calculations of fluctuation-induced
forces and torques. As described in the following 
section, our formalism computes the fluctuation-induced 
heating and force on a body essentially by averaging 
over a thermal ensemble of dipole sources distributed 
throughout the body, including at points arbitrarily
close to the surface---thus yielding the diverging
force contributions described above.
Physically, these contributions must all \textit{cancel}
when integrated over the body to compute the total force, 
and this cancellation is manifest in theoretical
approaches---such as scattering-matrix methods~\cite{Krueger2012}---
in which the integrals can be evaluated analytically.
However, any finite-resolution numerical code
will entail imperfect cancellation of divergent
force contributions from different surface regions,
yielding numerical noise that dominates the 
actual signal. For reasons discussed above, this
problem is \textit{not} present in numerical
heat-transfer solvers~\cite{Rodriguez2012, Polimeridis2015B}, 
which are thus
somewhat easier to implement; na\"ive attempts
to repurpose such a code by swapping out
Joule heating for Lorenz force in volume
integrals are doomed to failure (as we can
attest from experience). Instead, as we show
in Section IIIb, proper implementation requires
careful consideration of the physical requirements
of symmetry and reciprocity, which may be
exploited to reorganize force and torque calculations
in such a way as to achieve \textit{explicit} analytical
cancellation of the dangerous terms, yielding
an effective numerical method.
 
Meanwhile, all the complications we have noted thus far
arise already in determining the contribution of just
a single frequency to the NEQ force/torque; the indefinite
sign of momentum transfers creates additional headaches
when it comes to evaluating the frequency integrals
needed to compute total time averages of fluctuation-induced
quantities, which receive contributions from fluctuations 
on all time scales~\cite{Landau2}.
Indeed, as has been noted previously~\cite{Krueger2011, Krueger2012} 
and as we illustrate in the examples below (see especially
Figure \ref{GoldSpheresFigure}),
the net momentum emitted or absorbed by a body due
to frequency-$\omega$ fluctuations may oscillate rapidly
with $\omega$, with frequent sign changes and extensive
cancellations
that can only be resolved by evaluating large numbers
of costly integrand samples. This is again in contrast
to energy-transfer problems, where the frequency integrand
is of definite sign and no cancellations can arise~\cite{PolderVanHove1971}.
The upshot is to heighten the urgency of efficient algorithms
for computing the force/torque contributions of individual
frequencies, as numerical frequency quadratures can easily
require hundreds or thousands of samples even for 
convergence to one or two decimal places 
(Figure \ref{GoldSpheresFigure}).

If fluctuation-induced forces are so difficult to compute,
why was it possible for sophisticated computational tools
for \textit{equilibrium} Casimir forces to arise over the
last decade~\cite{Reynaud2006, Rahi2009, Reid2013B}---tools which, as noted above, predated
and laid the groundwork for the current generation of NEQ 
fluctuation-modeling tools~\cite{Lussange2012, Krueger2012, Rodriguez2012}?
One answer is that for equilibrium problems
there is a scalar quantity---the Casimir energy of a
configuration of bodies---that may be computed
and differentiated with respect to displacement
or rotation of the bodies to yield forces and torques~\cite{Milonni2013}.
This strategy is adopted explicitly by many
general-purpose equilibrium Casimir methods~\cite{Rahi2009, Johnson2011};
moreover, elsewhere we have rigorously demonstrated
that the alternative approach of surface-integrating 
the equilibrium thermal expectation value of the Maxwell 
stress tensor is exactly equivalent to differentiating 
an energy expression~\cite{Reid2013B}.
A second answer is that, at thermal equilibrium,
Wick rotation is available to convert oscillatory
real-frequency integrands into smoothly decaying
imaginary-frequency integrands of definite sign.~\cite{Landau2, Johnson2011}
This reduces the burden on numerical solvers 
for EQ problems by allowing accurate estimation of 
frequency integrals with many fewer integrand samples
than is typically required for NEQ force/torque problems.
Finally, the simple form of the force law for the
zero-temperature Casimir pressure between
infinite planar perfect mirrors~\cite{Casimir1948}
allows the use of the proximity-force approximation
(PFA)~\cite{Derjaguin1960} to estimate equilibrium
forces between closely-spaced bodies, while no such
simple approach is available for the NEQ case.

Although we are concerned in this paper with finite-size
bodies described by continuum material models, it is
interesting to note that the computational problems
we have discussed have analogues in the theory of
radiative losses from \textit{pointlike} 
particles~\cite{Moniz1977,Jackson1999,Landau2013, Yaghjian2008}.
For an accelerating pointlike particle, the classical
Larmor formula and its relativistic 
generalization~\cite{Jackson1999} give a straightforward
and quantitatively accurate description of the
\textit{power} loss, but the accompanying
\textit{force} (radiation reaction) on the particle 
is much more difficult to predict and---despite
more than a century of intensive investigation---remains
controversial to this day~\cite{Ford1991,Griffiths2010}.
It is possible that the new theoretical methods we
introduce in this paper could shed light on this 
longstanding problem; we will return to this question
in Section \ref{ConclusionsSection}.

\section{FVC/FSC trace formulas for non-equilibrium forces and torques }
\label{TheorySection}

The FVC and FSC methods respectively apply techniques
borrowed from
the volume-integral-equation (VIE)~\cite{Chew2009}
and surface-integral-equation (SIE)~\cite{Harrington1996}
approaches to classical scattering to compute rates of
fluctuation-induced energy and momentum transfer among
material bodies. Special cases of these formulas
for the specific case of energy-transfer problems were
presented in~\citeasnouns{Rodriguez2012}{Rodriguez2013B}
for the FSC case and in~\citeasnoun{Polimeridis2015B} for 
the FVC case; the objective of this section is to extend 
the FVC and FSC methods to the calculation of general
fluctuation-induced quantities---including powers, 
forces, and torques (PFTs)---and
to emphasize the shared conceptual foundations of the
two approaches, culminating in a single set of 
master formulas [\red{Equations} (\ref{MasterFormulas})]
expressing the predictions of both formalisms in a unified
language and notation.

The common physical picture underlying FVC and FSC methods
attributes fluctuation-induced phenomena to ``bare'' current
fluctuations---in the fluctuation-dissipation sense
of Rytov~\cite{Rytov1987}---which are ``dressed'' by the
polarization response of surrounding media, yielding
an effective fluctuation-induced source distribution
from which PFTs may be computed precisely as in classical,
deterministic scattering problems
(Section \ref{CommonTheorySection}). The result is
a family of formulas [\red{Equations} (\ref{MasterFormulas})]
expressing spectral densities of PFT contributions
as traces of products of matrices, with the various matrices
describing the distinct physical
ingredients outlined above (bare Rytov currents,
material polarization, and PFTs due to sources).
Although the particular forms assumed
by these matrices differ in the FVC and FSC cases
(Sections \ref{FVCTheorySection} and \ref{FSCTheorySection}),
one feature common to force/torque modeling in
\textit{both} frameworks is the need to organize
calculations in such a way as to realize a certain
analytical cancellation between particular contributions
(Section \ref{QPFTSection})---which, if not properly
handled, yield numerical noise that swamps the
correct results by several orders of magnitude.


\subsection{Common structure of FSC and FVC trace formulas}
\label{CommonTheorySection}

Among the various methods available for numerical solution
of electromagnetic scattering problems, integral-equation
(IE) methods are distinguished by their emphasis on the
\textit{sources} of the scattered fields: physical or
effective electric and magnetic currents induced
by incident fields in the bulk or surface of material 
bodies~\cite{Harrington1996, Chew2009}.
In contrast to other methods (such as the
finite-difference~\cite{Kunz1993}
or finite-element~\cite{Jin2014} methods) which solve 
directly for fields,
IE methods typically solve first for induced sources,
then compute the scattered fields and other physical
quantities of interest---including powers, forces and
torques (PFTs)---from the currents as a post-processing 
step~\cite{Reid2015B, Polimeridis2015A}.
More specifically, in a deterministic time-harmonic 
scattering problem at angular frequency $\omega$, 
let $\bmc C(\vb x)$ 
be the spatially-varying current distribution induced by 
incident fields impinging on a body; then the scattered
fields are obtained by a
linear convolution, $\bmc F \propto \bmc G \star \bmc C$,
and the time-average PFTs on the body---which are quadratic
functions of the fields---become quadratic forms
involving the currents~\cite{Reid2015B, Polimeridis2015A}:
\begin{subequations}
\begin{equation}
Q\supt{PFT} = \bmc C^* \star \bmc Q\supt{PFT} \star \bmc C
\qquad \text{(deterministic).}
\end{equation}
Here $\bmc F={\vb E \choose \vb H}$ is the field
six-vector, while the current six-vector 
$\bmc C={\vb J \choose \vb M}$
includes both electric currents $\vb J$ and magnetic currents $\vb M$
and may be either a volume current distribution existing throughout
the bulk of material bodies (for VIE methods~\cite{SWG1984})
 or a surface-tangential 
current distribution confined to the interfaces between bodies 
(for SIE methods~\cite{RWG1982}).
The convolution symbol $\star$ correspondingly represents 
integration over body volumes (VIE) or surfaces (SIE), 
and $\bmc G$ is the
$6\times 6$ dyadic Green's function (DGF)~\cite{Chew2009} of the vacuum (VIE)
or of the homogeneous medium in which we are computing
the field (SIE).
The convolution kernel 
$\bmc Q\supt{PFT}$ depends on the formulation and on the quantity
being computed; specific expressions are given 
below. Equation (\ref{CQC}a) omits a possible incident-field term
that may be present in some scattering problems but is absent
for the fluctuational problems considered here; this, and the 
structure of the $\bmc Q\supt{PFT}$ operator in SIE and VIE 
formulations, are discussed briefly in 
Section \ref{QPFTSection} below and in more detail
in~\citeasnoun{Reid2015B}.

Numerical IE solvers discretize by approximating the unknown
current distribution as an expansion in some convenient
discrete set of basis functions,
\numeq{CExpansion}
{\bmc C(\vb x)\approx \sum_{\alpha=1}^{N\subt{BF}}
  c_\alpha \bmc B_\alpha(\vb x),}
where the basis functions $\{\bmc B_\alpha\}$ describe
volume currents~\cite{SWG1984} (VIE) or surface currents~\cite{RWG1982}
(SIE). Scattering problems reduce to solving for the 
$N\subt{BF}$-dimensional
vector $\vb c$ of current coefficients~\cite{Chew2009}, after which
time-average PFT quantities are computed as vector-matrix-vector
products~\cite{Reid2015A},
\begin{equation}
Q\supt{PFT}=\vb c^\dagger \vb Q\supt{PFT} \vb c 
=\text{Tr }\Big[ \vb Q\supt{PFT} \vb c \vb c^\dagger \Big]
 \,\, \text{(deterministic)}
\end{equation}
where $\vb Q\supt{PFT}$ is the $N\subt{BF}\times N\subt{BF}$ matrix
of the operator $\bmc Q\supt{PFT}$ in the $\{\bmc B_\alpha\}$ basis.
(For reasons that
will be clear shortly, we have here rewritten the 
vector-matrix-vector product as the trace of a matrix
product involving the rank-one outer-product matrix
$\vb c\vb c^\dagger$.) 

The emphasis on sources makes IE methods particularly
well-suited to the study of Casimir forces and other
fluctuation-induced phenomena, which arise from
thermal and quantum-mechanical fluctuations of the
currents $\bmc C$ in material bodies~\cite{Landau2}. 
The stochastic
nature of these currents ensures a vanishing mean value 
$\ExpVal{\bmc C}=0$ for any individual current component,
but \textit{products} of current components in lossy
bodies may have nonvanishing thermal averages,
prescribed by fluctuation-dissipation theorem
in the form of Rytov~\cite{Rytov1987}:
\begin{align}
 \Big< \mc C_\mu\sups{free}(&\vb x) \, 
       \mc C_\nu\sups{free*}(\vb x^\prime)
 \Big>_{\omega}
 =\mc R_{\mu\nu}(\vb x) \delta(\vb x-\vb x^\prime),
\\
\bmc R(\vb x)
&\equiv
  \frac{2k\bmc Z_0^{-1}}{\pi} 
  \Theta\Big(T(\vb x), \omega\Big) \text{Im }\bmc X(\vb x),
\nn
\Theta(T,\omega)
&=
   \hbar \omega \left[ \frac{1}{e^{\hbar\omega/kT}-1} + \frac{1}{2}\right],
\nonumber
\end{align}
where
$\bmc X=\left(\begin{smallmatrix}
        \boldsymbol{\epsilon}-\vb 1 & 0 \\ 0 & \vbMu-\vb 1
        \end{smallmatrix}\right)
$
is the spatially-varying material susceptibility tensor
and 
$\bmc Z_0= \left(\begin{smallmatrix}
  Z_0 & 0 \\ 0 & Z_0^{-1}
  \end{smallmatrix}\right)
$
with $Z_0\approx 377\,\Omega$ the impedance of vacuum.
\red{[Here and below, the symbols ``Re'' and ``Im''
applied to matrices denote Hermitian symmetrizing
and anti-symmetrizing \textit{matrix} (not elementwise)
operations.]}
Thus quadratic functions of currents---such
as the time-average PFT quantities in (\ref{CQC}a) 
and (\ref{CQC}c)---may have nonzero thermal 
averages, whose values are readily computed in the 
IE framework.

The superscript on $\mc C$ in (\ref{CQC}d) indicates
that the right-hand side of that equation is to be interpreted
as the mean-square amplitude of a ``free'' current distribution
$\bmc C\sups{free}(\vb x)$ embedded in the surface or volume of
a material body. The \textit{total} current contributing to
PFTs in (\ref{CQC}a) includes both $\bmc C\sups{free}$
and the \textit{induced} current $\bmc C\sups{ind}$
resulting from the polarization response of the body to
$\bmc C\sups{free}$; in linear media---our
exclusive focus in this paper---the total current
may be represented as a linear convolution of the
form~\cite{Rytov1987,Landau3}
\numeq{CCC}
{
\bmc C=\bmc C\sups{free} + \bmc C\sups{ind} = \bmc W \star \bmc C\sups{free}
}
where we have introduced the symbol $\bmc W$ to denote a certain
linear operator---depending through $\bmc X$ on the material geometry
and discussed in more detail below---which
``dresses'' the bare distribution $\bmc C\sups{free}$
to yield the total current including the polarization response 
of the material bodies.
For example, at low frequencies the $\bmc W$ operator for a
homogeneous dielectric sphere reduces to a simple 
multiplicative factor, 
$\mc W\sups{$\epsilon$-sphere}\xrightarrow{\omega \to 0}\frac{3}{(2+\epsilon)}$.

Using $\bmc W$ to ``dress'' the bare Rytov source distribution
(\ref{CQC}d) now yields the correlation function of the full current
distribution in (\ref{CQC}a),
\begin{align}
 \ExpVal{\bmc C \bmc C^*}_\omega 
 &= \bmc W \star \bmc R \star \bmc W^\dagger,
\nonumber
\intertext{or, in terms of the discretized representation (\ref{CQC}c),}
 \ExpVal{\vb c \vb c^\dagger}_\omega 
&= \underbrace{\vb W \vb R \vb W^\dagger}_{\vb D}
\intertext{where $\vb R$ and $\vb W$ are the matrix representations of
           $\bmc R$ and $\bmc W$ and convolutions have become
           matrix multiplications; we refer to $\vb R$ and $\vb D$
           here as the ``Rytov'' and ``dressed Rytov'' matrices,
           with the latter describing the bare Rytov distribution
           as ``dressed'' by the polarization response of the material
           geometry.
           Then the thermally-averaged PFT (\ref{CQC}c) reads}
 \ExpVal{Q\supt{PFT}}_\omega 
 &= \Tr\Big[ \vb Q\supt{PFT} 
    \ExpVal{\vb c \vb c^\dagger}_\omega
        \Big]
\nn
 &= \Tr\Big[ \vb Q\supt{PFT} \, 
       \vb W  \vb R \vb W^\dagger \Big]
    \quad\text{(fluctuational}).
\end{align}
Equation (\ref{CQC}h) is the fluctuational analogue
of the deterministic formula (\ref{CQC}c); in particular, the matrix
$\vb Q\supt{PFT}$ is the \textit{same} matrix in both cases,
and the formulas differ only in that the rank-one matrix
$\vb c \vb c^\dagger$ describing deterministic currents in
(\ref{CQC}c) is replaced by the
\textit{full}-rank matrix $\vb D=\vb W\vb R\vb W^\dagger$
describing the effective mean-square distribution
of fluctuation-induced currents contributing to PFTs.

Equation (\ref{CQC}g) gives the spectral density of PFT contributions
arising from source fluctuations at frequency $\omega$; the full
thermally-averaged PFT then follows by accounting for the contributions
of all frequencies,
\begin{equation}
 \ExpVal{Q\supt{PFT}}=\int_0^\infty \ExpVal{Q\supt{PFT}}_\omega \,d\omega.
\end{equation}
\label{CQC}%
\end{subequations}
The logical sequence of physical ideas 
leading from (\ref{CQC}a) to (\ref{CQC}g)
goes through equally well in both VIE and SIE frameworks
and for any choice of basis functions; all that differs are the
particular forms assumed by the $\vb Q, \vb R,$ and $\vb W$ 
matrices.
In following two sections we discuss the
computation of the $\vb R$ and $\vb W$ matrices
in the FVC and FSC frameworks, while in Section \ref{QPFTSection}
we discuss computation of the $\vb Q$ matrices
in both frameworks.

\subsection{Non-equilibrium forces and torques in the VIE context: The FVC force and torque formulas}
\label{FVCTheorySection}

In VIE solvers~\cite{Chew2009, Volakis2012},
$\bmc C={\vb J \choose \vb M}$
is a six-vector volume current distribution whose
magnetic components $\vb M$ vanish for non-magnetic
media $(\mu=1)$.
The induced portion of this current, $\bmc C\sups{ind}(\vb x)$,
is typically nonzero at any point $\vb x$ inside a material
body, where it is linearly related to the total electromagnetic
fields at $\vb x$ through a $6\times 6$ susceptibility tensor
$\bmc X$~\cite{Chew1995}:
\numeq{CFromFTot}
{
 \bmc C\sups{ind}(\vb x)
 = -ik \bmc Z_0^{-1} \bmc X(\vb x) \bmc F \sups{tot}(\vb x)
}
$$ k=\frac{\omega}{c},
   \qquad
   \bmc X=
    \left(\begin{smallmatrix}
    \boldsymbol{\epsilon}-\vb 1 & 0 \\ 0 & \vbMu-\vb 1
   \end{smallmatrix}\right),
   \quad 
   \bmc Z_0\equiv 
                 \left( \begin{smallmatrix} Z_0 & 0 \\ 
                                            0 & Z_0^{-1}
                        \end{smallmatrix}
                 \right)
$$
[For non-magnetic media ($\vbMu=\vb 1)$ this reduces to simply
$\vb J=-i\omega\epsilon_0(\boldsymbol{\epsilon}-\vb 1)\vb E\sups{tot}$
\red{with $\vb 1$ the $3\times 3$ identity matrix.}]
To derive an expression for the dressing matrix $\vb W$,
we take $\bmc F\sups{tot}$ to be the
field radiated by the bare current $\bmc C\sups{free}$
plus the field of the corresponding induced 
current, i.e. 
\numeq{CIndFromCIndCFree}
{
 \bmc C\sups{ind}
=k^2 \bmc X\Big[\bmc G^0\star
 \big(\bmc C\sups{free} + \bmc C\sups{ind}\big)\Big]
}
with $\bmc G^0$ the $6\times 6$ vacuum Green's
dyadic,
related to the usual $3\times 3$ and scalar Green's 
functions~\cite{Collin1991, Tai1994, Scheel2009} by
\begin{subequations}
\begin{align}
\bmc G^0
&\equiv
   \left(\begin{array}{cc} \hphantom{-} \mb G^0 & \mb C^0 \\
                                       -\mb C^0 & \mb G^0
   \end{array}\right)
\\
 \mb G^0_{ij}(\vb r)
&= \left[\delta_{ij} +\frac{1}{k^2} \partial_i \partial_j\right]
   \frac{e^{ik|\vb r|}}{4\pi|\vb r|}
\\
&= \frac{e^{ikr}}{4\pi k^2 r^3}
   \Big[ f_1(ikr) \delta_{ij} + f_2(ikr) \frac{r_i r_j}{r^2} \Big],
\\[5pt]
  \mb C^0_{ij}(\vb r)
&= -\frac{1}{ik} \varepsilon_{ijk} \partial_k \mb G^0_{k\ell}
\\
&=\frac{e^{ikr}}{4\pi (ik) r^3} f_3(ikr) \varepsilon_{ijk} r_k,
\end{align}
\label{GCDef}%
\end{subequations}
$$
  f_1(x) \equiv -1 + x -x^2,   \,\,\,
  f_2(x) \equiv  3 - 3x + x^2, \,\,\,
  f_3(x) \equiv -1 + x.
$$
\red{Using (\ref{CIndFromCIndCFree}) to compute the total current
$\bmc C=\bmc C\sups{free}+\bmc C\sups{ind}$ 
yields an expression for the ``dressing'' operator $\bmc W$
in (\ref{CCC}):}
$$
 \bmc W \equiv \vb 1 + \big(\vb 1 - k^2 \bmc X \bmc G^0\big)^{-1}.
$$
For a given basis of expansion functions $\{\bmc B_\alpha\}$,
\red{the matrix of the $\bmc W$ operator may be 
related to the matrices of the $\bmc X$ and $\bmc G^0$ operators} according to
\numeq{WFromVG}
{ \vb W = \Big[\vb 1 + \vb S^{-1} \vb V \vb S^{-1} \vb G^0\Big]^{-1} }
where the various matrices have elements
\begin{subequations}
\begin{align}
S_{\alpha\beta}
&=\Inp{\bmc B_\alpha}{\bmc B_\beta}, 
\\
V_{\alpha\beta}
&=-k^2\Exptwo{\bmc B_\alpha}{\bmc X}{\bmc B_\beta},
\\
G^0_{\alpha\beta}
&=\Exptwo{\bmc B_\alpha}{\bmc G^0}{\bmc B_\beta}.
\end{align}
(The notation $\vb V$ for the matrix of the quantity $-k^2\bmc X$
follows~\citeasnouns{Rahi2009}{Krueger2012}, where this quantity
is interpreted as a scattering-theoretic ``potential.'')
\red{Similarly, the matrix of the Rytov operator $\bmc R$
[Equation (\ref{CQC}d)] describing the fluctuation-induced
free source distribution dressed by $\bmc W$ has elements
\numeq{RMatrixElements}
{ R_{\alpha\beta}
 =\frac{2k}{\pi} 
  \Exptwo{\bmc B_\alpha}{\bmc Z_0^{-1}\,\Theta\,\text{Im }\bmc X}
         {\bmc B_\beta}.
}}
\label{MatrixElements}%
\end{subequations}
The task of a
numerical~\cite{buff-em} FVC solver
is then simply to evaluate, \red{for an appropriate
set of volume-current basis functions $\{\bmc B_\alpha \}$,
the numerical integrals~\cite{Reid2016D} defining the matrix elements 
(\ref{MatrixElements}),
then evaluate the matrix-product trace in
(\ref{CQC}g) via standard methods of numerical
linear algebra~\cite{Lapack}.}

\red{Equation (\ref{WFromVG}) may be derived from equation
 (\ref{CIndFromCIndCFree}) by expanding $\bmc C\sups{ind}$ and
$\bmc C\sups{free}$ via (\ref{CExpansion}) and using the
completeness relation~\cite{Boyd2001}
\numeq{Completeness}
{ \sum_{\alpha \beta}
   S^{-1}_{\alpha\beta} \,
   \mc B^*_{\alpha\mu}(\vb x) 
   \mc B_{\beta \nu}(\vb x^\prime)
   = \delta_{\mu\nu}\delta(\vb x-\vb x^\prime).
}
For a finite basis set---such as the SWG~\cite{SWG1984} functions
used for the FVC computations of Section \ref{PhotonTorpedoSection}
in which the number of basis functions is proportional to
the number of tetrahedra
into which objects are discretized---equation (\ref{Completeness}) is 
only an approximate equality, with the approximation improving as the
basis set is refined~\cite{Boyd2001}.}

\subsection{Non-equilibrium forces and torques in the SIE context: The FSC force and torque formulas}
\label{FSCTheorySection}

In SIE solvers, $\bmc C(\vb x)={\vb K \choose \vb N}$ is an
effective tangential
\textit{surface} current distribution localized on interfaces between
homogeneous material regions.
Although the basic logical sequence leading to equation (\ref{CQC}g)
remains unchanged in this case---PFT quantities are quadratic 
functions of
currents, thermal averages of current-current products are
related to temperatures and susceptibilities by
fluctuation-dissipation relations, and assembling these
ingredients yields trace formulae of the form (\ref{CQC}g)---the
physical interpretation of $\bmc C$ is now subtly different. One
immediate clue is the fact that the magnetic surface current
$\vb N$ is generally nonzero even for non-magnetic media. As
this observation demonstrates, the surface currents in SIE 
solvers do not coincide with the physical sources of the
scattered fields; instead, $\bmc C$ is properly understood 
as a tally of boundary data recording the total tangential
$\vb E$ and $\vb H$ fields present at material 
interfaces~\cite{StrattonChu, Medgyesi1994}.

Despite this shift in viewpoint, it remains 
straightforward to write SIE versions of equations like
(\ref{CQC}a) and (\ref{CQC}c) expressing PFT quantities
in classical deterministic problems
as quadratic functions of the effective surface currents~\cite{Reid2015B}.
On the other hand, the derivation of appropriate
bare and dressed Rytov matrices
$\vb R$ and $\vb D$
for use in (\ref{CQC}g) is more subtle,
as it is not obvious how one translates the Rytov correlation
function (\ref{CQC}1d) for physical volume 
currents into a statement about effective surface
currents.
Nonetheless, as shown in~\citeasnoun{Rodriguez2013B}, in the
SIE context it remains possible to write down a 
$\vb D$ matrix appropriate for use in
equation (\ref{CQC}g): for PFTs induced by sources in body $s$
among a collection of one or more homogeneous bodies,
the matrix
\numeq{DSIE}
{ \vb D = 
  -\frac{4}{\pi}
  \Theta(T_s, \omega) \vb M^{-1} \Big( \text{sym }\vb G^s \Big) 
                      \vb M^{-1\dagger}
}
plays a role equivalent to that of $\vb W\vb R \vb W^\dagger$
in (\ref{CQC}g).
Here $T_s$ is the temperature of the source body (assumed constant
throughout the body) and 
$\vb M$ is the discretized SIE matrix for the collection
of objects (which enters classical scattering problems
through a relation of the form $\vb M \vb c = \vb v$ with
$\vb v$ the basis-set projection of the incident field);
for a single homogeneous body, $\vb M$
is just the sum of the matrix representations, in the
$\{\bmc B_\alpha\}$ basis,
of the Green's dyadics for the material media
exterior and interior to the body,
$ M_{\alpha\beta} = 
  \Exptwo{\bmc B_\alpha}{\bmc G\sups{ext} + \bmc G\sups{int}}
         {\bmc B_\beta}.
$~\cite{Medgyesi1994}.
The matrix $\left(\text{sym }\vb G^s\right)$ in 
(\ref{DSIE}) is a symmetrized version of the portion of $\vb M$
that describes the self-interactions of surface currents on 
body $s$, retaining only the contributions of the medium
interior to that body.

The structure of equation (\ref{DSIE}) is similar to that
of (\ref{CQC}f); in both cases we have a matrix
describing \textit{bare} source fluctuations
($\vb R$ or $\text{sym } \vb G^s$)
that is conjugated (``dressed'') by a matrix describing
the polarization response of surrounding media
($\vb W$ or $\vb M^{-1}$) to yield a matrix ($\vb D$) 
describing the full (physical volume or
effective surface) thermal current distribution,
which may be paired with various $\vb Q\supt{PFT}$
matrices to compute various fluctuation-induced
PFT quantities.

\red{However, a key difference between SIE and VIE is that
the VIE formulation for material bodies
in vacuum or a homogeneous, isotropic background medium
refers only to the dyadic Green's function $\bmc G^0$
of the background medium,
for which closed-form analytic expressions are
available; this is true \textit{irrespective} of the
complexity of the material properties of the bodies,
which may---for example---be anisotropic or continuously
spatially varying without requiring any modification
of the VIE solution procedure.
In contrast, the SIE formulation requires knowledge of the
Green's functions for the media both exterior
\textit{and interior} to the scattering bodies; 
in practice, this restricts FSC methods
to problems involving only bodies of homogeneous,
isotropic permittivity and permeability
and homogeneous temperature.
For VIE there is no such limitation, so 
the FVC method is the appropriate choice for
bodies with inhomogeneous temperature distributions
or continuously varying or anisotropic
material properties.}

\subsection{Methods for writing $\vb Q\sups{PFT}$ matrices:
            Analytical cancellation of numerical divergences}
\label{QPFTSection}

The PFT trace formula (\ref{CQC}f) involves two ingredients:
the dressed Rytov matrix $\vb D$, discussed in the previous 
two sections, which describes thermal current fluctuations as
dressed by the electromagnetic
response of surrounding media, and the $\vb Q\supt{PFT}$ matrix,
describing the power, force, and torque on bodies as a
quadratic function of the currents.
Whereas the derivation
of the $\vb D$ matrix relies on quantum-statistical-mechanical
reasoning as embedded in the Rytov correlation function
(\ref{CQC}d), the $\vb Q$ matrix involves strictly
classical, deterministic considerations.
Elsewhere we have
presented detailed discussions of the structure of the classical
$\vb Q$ matrices in the IE~\cite{Reid2015B} framework;
as discussed in that reference,
for each PFT quantity there are in fact multiple \textit{distinct}
methods for writing $\vb Q$ matrices, all equivalent in
exact arithmetic and with complete basis sets, but differing
in accuracy and efficiency for practical numerical computations.
In many classical scattering problems, the various approaches are
somewhat interchangeable, offering similar computational cost and
accuracy; for fluctuational problems, on the other hand, one
particular strategy emerges as the clear favorite, as we now
discuss.

To compute the absorbed power or force on a body, one may
evaluate either \textbf{(a)} a surface
integral~\cite{BohrenHuffman1983, Mishchenko2002}, involving
the Poynting vector (PV) or Maxwell stress tensor (MST),
over the body surface or over a closed bounding surface 
surrounding but displaced from the body
[the ``displaced-surface-integral PFT'' (DSIPFT) approach], or
\textbf{(b)} a volume integral~\cite{Krueger2012},
involving the local
Joule heating $\frac{1}{2}\text{Re }\bmc C^\dagger \bmc F$
or Lorentz force
$\frac{1}{2\omega}\text{Im }\bmc C^\dagger \nabla \bmc F$
inside the body.
The latter case further bifurcates according as
we relate $\bmc C$ to $\bmc F$ using 
equation (\ref{CFromFTot})---the ``overlap PFT'' (OPFT) approach,
so called because the $\vb Q$ matrix elements in this case
involve overlap integrals between basis functions---or using the 
convolution $\bmc F=\bmc G \star \bmc C$, 
which we term the ``energy-momentum transfer PFT'' (EMTPFT) approach
because PFTs in this case may be interpreted
as transfers of energy or momentum from incident fields
to currents~\cite{Reid2015B}.
(In the SIE case, the OPFT may alternatively be derived by
collapsing the bounding surface of the DSIPFT approach
to the body surface and expressing the PV or MST
in terms of surface currents~\cite{Reid2015B}.)

In principle we thus have three distinct algorithms---DSIPFT, 
EMTPFT, or OPFT---for computing $\vb Q\sups{PFT}$ matrices.
However, the OPFT approach---while computationally
the most efficient method, and one which generally yields
acceptable accuracy in deterministic problems---turns out
to be unsuitable for force and torque calculations
in fluctuational settings, for reasons discussed
in~\citeasnoun{Reid2015B}.
On the other hand,
the DSIPFT approach typically yields good accuracy
in both deterministic and fluctuational problems,
but---though cost-competitive with other methods
in the deterministic case---is prohibitively
expensive for fluctuational calculations. Briefly,
the reason is that computing each element
of the $\vb Q$ matrix involves a costly individual 
numerical cubature over the bounding surface; for
the rank-one calculation of equation (\ref{CQC}c)
the calculation may be rearranged to avoid explicitly
forming $\vb Q$, but for the full-rank problem
of equation (\ref{CQC}g) there is no bypassing
the time-consuming evaluation of $\vb Q$ elements.

This leaves the EMTPFT strategy as the only viable
option for accurate and efficient evaluation
of fluctuation-induced forces and torques in
both the SIE and VIE frameworks. Here we briefly
sketch the essentials of this approach. 

In EMTPFT we identify the power absorption, force, and
torque on a body with the energy, linear momentum, and
angular momentum transferred to currents in the body---either
physical volume currents in the VIE case, or effective 
surface currents in the SIE case---by the ambient fields.
Thus, in a VIE solver 
\red{the power~\cite{Polimeridis2015A, Polimeridis2015B},}
force and torque on a body may be computed as volume integrals 
of the form
\begin{subequations}
\begin{align}
  \vb P
&=\frac{1}{2}\text{Re }\int_{\mc V} \bmc C^* \bmc F \, dV
\\
  \vb F
&=\frac{1}{2\omega}\text{Im }\int_{\mc V} \bmc C^* \nabla \bmc F \, dV
\\
 \bmc T
&=
  \frac{1}{2\omega}\text{Im }\int_{\mc V} 
  \underbrace{
  \big\{ \bmc C^* \times \bmc F + \bmc C^*(\vb r\times \nabla) \bmc F
             }_{\bmc C^* \wt{\nabla}\bmc F }
  \big\} \, dV
\end{align}
\label{FTVI}%
\end{subequations}
\red{The symbol $\wt{\nabla}$ for the operator defined by (\ref{FTVI}c)
is a convenient shorthand notation for torque calculations.}

\red{The VIE power formula (\ref{FTVI}a) was 
discussed extensively in the classical scattering context 
in~\citeasnoun{Polimeridis2015A}
and in the radiative heat-transfer context in~\citeasnoun{Polimeridis2015B};
the latter reference also quoted the force formula (\ref{FTVI}b),
but omitted details.}
Equations (\ref{FTVI}b,c) are intuitively reminiscent of the expressions
$\vb p \cdot \nabla \vb E$ and $\vb p \times \vb E$ for the force and
torque on a point dipole in an external electrostatic field~\cite{Jackson1999}; 
they
may be derived rigorously by considering the usual Lorentz force
on the charges and currents in an infinitesimal volume and applying
Maxwell's equations and Stokes' theorem to the volume integral.

When $\bmc F$ in (\ref{FTVI}) is the field radiated by $\bmc C$
itself---a situation encountered when computing the self-force
or self-torque on a body due to thermal fluctuations within it---the 
force reads
\numeq{ForceIntegral}
{
 F_i  = \frac{1}{2c}\text{Im}
      \iint\!\!
      {\vb J \choose \vb M}^\dagger\!\!
      \left(\begin{array}{cc}
       iZ_0 \partial_i \mb G^0 & i\partial_i \mb C^0 \\
       -i \partial_i \mb C^0   & iZ_0^{-1} \partial_i \mb G^0\!\!
      \end{array}\right)\!\!
      {\vb J \choose \vb M}
      \, d^2V 
}
\red{with $\mb G^0, \mb C^0$ the dyadic Green's functions
of vacuum (or the embedding medium), Equation (\ref{GCDef}).}
Although equation (\ref{ForceIntegral}) would appear to define a 
\textit{singular} integral in view of the short-distance singularities 
of the kernels (\ref{GCDef}) 
[e.g. $\mb G^0(\vb r)\sim \frac{1}{|\vb r|^3}$ as 
$|\vb r|\to 0$], this appearance is misleading; upon rearranging 
and exploiting the reciprocity relations
$\{\mb G^0, \mb C^0\}_{ij}(\vb r) = \{\mb G^0, \mb C^0\}_{ji}(-\vb r)$
one finds that singular terms in the
integrand explicitly cancel in (\ref{ForceIntegral}),
yielding an expression of the form
\begin{align}
 F_i  &= \frac{1}{2c} \iint
\Bigg\{
  \text{Im } \Big(Z_0 J_j^* J_k + Z_0^{-1} M_j^* M_k\Big)
  \text{Im } \partial_i \mb G^0_{jk}
\nn 
&\hspace{0.5in}
 -
  \text{Re } \Big(J_j^* M_k - M_j^* J_k\Big)
  \text{Re } \partial_i \mb C^0_{jk}
\Bigg\}d^2V
\label{NonsingularForceIntegral}
\end{align}
involving only the nonsingular operators 
$\text{Im } \mb G^0$ and $\text{Re } \mb C^0$, which behave for 
small $\vb r$ like~\cite{Tai1994, Scheel2009}
\begin{subequations}
\begin{align}
\text{Im }\mb G^0_{ij}(\vb r)
&= \frac{k}{6\pi}\delta_{ij}
  -\frac{k^3 r^2}{30\pi}
   \Big[\delta_{ij} - \frac{1}{2}\frac{r_i r_j}{r^2}\Big]
  +O(r^4)
\\
\text{Re }\mb C^0_{ij}(\vb r)
&= \left[\frac{-k^2}{12\pi} + \frac{k^4 r^2}{120\pi} + O(r^4)\right]
     \varepsilon_{ijk} r_k.
\end{align}
\label{GCLimits}%
\end{subequations}
\red{Note that our derivation of (\ref{NonsingularForceIntegral})
used only the reciprocity of the \textit{exterior} medium, not that
of the scatterer, and thus remains valid for non-reciprocal
scatterers.}

\red{Analogous manipulations of the power and torque formulas
(\ref{FTVI}a,c) show that the contributions of singular
operator terms cancel from these integrals as well;
discretizing according to (\ref{CExpansion}) then yields
trace formulas (\ref{CQC}c) for the power and $i$-directed
force and torque with elements of the $\vb Q\supt{P,F,T}$ matrices 
involving inner 
products of volume-current basis functions with the nonsingular operators
$\text{Im }\mb G^0$, $\text{Re }\mb C^0$ and their derivatives:
\begin{subequations}
\begin{align}
 Q\supt{P}_{\alpha\beta}
 &=\frac{\omega}{2c}
  \EXPTWO{\bmc B_\alpha}{
  \left(\begin{array}{cc}
   Z_0 \text{Im } \mb G^0 & -\text{Re } \mb C^0 \\
  +    \text{Re } \mb C^0 & +Z_0^{-1}\text{Im } \mb G^0
        \end{array}\right)}{\bmc B_\beta},
\\
 Q\supt{F$_i$}_{\alpha\beta}
 &=\frac{1}{2c}
  \EXPTWO{\bmc B_\alpha}{
  \left(\begin{array}{cc}
   Z_0 \text{Im }\partial_i \mb G^0 & -\text{Re }\partial_i \mb C^0 \\
  +    \text{Re }\partial_i \mb C^0 & +Z_0^{-1}\text{Im }\partial_i \mb G^0
        \end{array}\right)}{\bmc B_\beta},
\\
 Q\supt{T$_i$}_{\alpha\beta}
 &=\frac{1}{2c}
  \EXPTWO{\bmc B_\alpha}{
  \left(\begin{array}{cc} 
        Z_0 \text{Im }\wt{\partial}_i \mb G^0 &
           -\text{Re }\wt{\partial}_i \mb C^0 \\
           +\text{Re }\wt{\partial}_i \mb C^0 & 
   +Z_0^{-1}\text{Im }\wt{\partial}_i \mb G^0
        \end{array}\right)}{\bmc B_\beta}
\end{align} 
\label{JDEPFTQMatrices}%
\end{subequations}}
\red{
Here $\wt{\partial_i}$ denotes the $i$th component of the 
operator $\wt{\nabla}$ defined by (\ref{FTVI}c).}

Equations (\ref{FTVI}) and (\ref{JDEPFTQMatrices}) are for the
VIE case. The SIE case is closely analogous, with
\textbf{(a)} the volume integrals in (\ref{FTVI}) replaced by
surface integrals;
\textbf{(b)} the current $\bmc C$ now understood to represent
effective (equivalence-principle) electric and magnetic
surface currents instead of physical volume currents;
\textbf{(c)} the matrix elements (\ref{JDEPFTQMatrices})
now involving 4-dimensional integrals over surface basis
functions instead of 6-dimensional integrals over volume basis
functions; 
\textbf{(d)} the vacuum dyadics $\mb G^0, \mb C^0$ replaced
by the dyadic Green's functions for the
homogeneous medium exterior to the body (if it is not
vacuum). These retain the nonsingular limiting forms
(\ref{GCLimits}) \textit{as long as the exterior medium
is lossless} (so that $k$ is real-valued), but
exhibit short-distance singularities in lossy exterior
media that invalidate EMTPFT force and torque
(but not power~\cite{Rodriguez2013B})
calculations in that case---a state of affairs
that is hardly unexpected given the murky
status of electromagnetic momentum 
(but not power~\cite{Tai2000}) in lossy
media~\cite{Landau3}.

The explicit reciprocity-enabled cancellation of the singular
contributions to (\ref{ForceIntegral}) affords a major reduction
in the computational cost of EMTPFT calculations, as the integrals
(\ref{JDEPFTQMatrices}) defining
$\vb Q$-matrix elements---whose computation typically dominates
the cost---are now \textit{nonsingular} and may be
evaluated by simple numerical cubature~\cite{Cools2003},
obviating the
need for complicated and expensive techniques for
evaluating singular integrals~\cite{Reid2015B, Reid2016D}.
For \textit{power} computations, this speedup is convenient
but inessential; the singular contributions to
$\vb Q\supt{P}$-matrix elements cancel harmlessly out of
numerical evaluations of (\ref{CQC}g), and their retention
degrades efficiency but not accuracy.

For \textit{force} and \textit{torque} computations,
on the other hand, the cancellation of singular
contributions effected by the transition from
(\ref{ForceIntegral}) to (\ref{NonsingularForceIntegral})
is not just a computational convenience, but a
crucial ingredient in establishing a
tractable numerical algorithm.
The difficulty is that the most singular contributions
to (\ref{ForceIntegral})---specifically, the
contributions associated with the $\frac{1}{|\vb r|^3}$
terms in $\mb G_0(\vb r),\mb C_0(\vb r)$---describe
force contributions that are essentially electrostatic
in nature, and the integral (\ref{ForceIntegral})
effectively sums force contributions between pairs of
electrostatic dipoles separated by arbitrarily short
distances $d$, yielding force contributions diverging
like $\sim d^{-4}$~\cite{Jackson1999}. 
Physically these contributions must all cancel,
and this cancellation occurs naturally when
the integral is evaluated analytically as is done in 
scattering-matrix approaches~\cite{Krueger2012}.
But na\"ive 
discretization of (\ref{ForceIntegral}) using
localized basis functions---such as the piecewise-linear
tetrahedron-based SWG functions~\cite{SWG1984}
or triangle-based RWG functions~\cite{RWG1982}
we use below (Sections \ref{ValidationSection},
\ref{ApplicationsSection})---destroys
the exact cancellation, resulting in
overwhelming numerical noise swamping the
signal in equation (\ref{ForceIntegral}).
(We emphasize again that this disease does not
afflict power computations, as there is no
power exchange between electrostatic dipoles~\cite{Jackson1999}.)
Thus the exact cancellation that converts
(\ref{ForceIntegral}) into (\ref{NonsingularForceIntegral})
is not merely an optional computational acceleration,
but rather an essential ingredient required for
any tractable numerical implementation of the FSC/FVC
approach to non-equilibrium force and torque computation.

\subsection{Master FSC/FVC formulas for non-equilibrium PFTs}
\label{MasterFormulasSection}

\red{
Assembling the ingredients from the preceding sections,
we obtain the following master formulas for the spectral density
at frequency $\omega$ of contributions
to powers, forces, and torques in both FSC
and FVC contexts:
\begin{subequations}
\begin{align}
\ExpValb{P}_\omega
&=
 \text{Tr }
 \Big[  \vb Q\supt{P}
        \,\text{Re }\big( \vb W \vb R \vb W^\dagger \big) 
 \Big]
\\
\ExpValb{F_i}_\omega
&= \text{Tr }
   \Big[ \vb Q\supt{F$_i$}
         \text{Im }\big(\vb W \vb R \vb W^\dagger\big) 
   \Big]
\\
\ExpValb{\mc T_i}_\omega
&=
   \text{Tr }\Big[ \vb Q\supt{T$_i$}
                   \text{Im }\big( \vb W \vb R \vb W^\dagger \big)
             \Big]
\end{align}
\label{MasterFormulas}
\end{subequations}
As discussed above, the $\vb Q$ matrices in these formulas---whose
entries are inner products of surface- or volume-current 
basis functions with
desingularized dyadic Green's functions and their
derivatives, Eq. (\ref{JDEPFTQMatrices})---describe the
PFT contributions of pairs of unit-amplitude basis functions,
while the dressed Rytov matrices $\vb D=\vb W\vb R \vb W^\dagger$
describe the mean-square amplitudes of fluctuation-induced currents
as dressed by the polarization response of the material
geometry.
In the following sections we use SIE and VIE implementations of 
these formulas to predict new nonequilibrium force and torque 
phenomena in nanoparticle sytems.
}

\red{
The fact that the power formula in (\ref{MasterFormulas}) involves
$\text{Re } \vb D$, while the force and torque formulas 
involve
$\text{Im } \vb D$, may be traced back to \textbf{(a)}
the \text{Re } and \text{Im } operators in the volume-integral
expressions for power and for force/torque in Eq. (\ref{FTVI}),
together with \textbf{(b)} the fact that the $\vb Q\supt{P}$
matrix in (\ref{JDEPFTQMatrices}) is symmetric, while 
$\vb Q\supt{F}$ and $\vb Q\supt{T}$ are antisymmetric.
}

\red{
The \textit{dimension} of the matrix traces in (\ref{MasterFormulas})
is the total number of basis functions used to represent volume or
surface currents in the material geometry; 
for the discretized-mesh-conforming
localized basis functions we choose here (discussed in the
following section), the number of basis functions required
to achieve reasonable accuracy
is on the order of a few hundred to a few thousand.
}

\subsection{Choice of basis functions}
\label{BasisFunctionSection}

The FVC and FSC formulas presented above are agnostic
with respect to the choice of basis functions
$\bmc B_\alpha(\vb x)$ used to represent current 
distributions [equation (\ref{CQC}b)]. In practice,
many choices of basis function are available, with
the appropriate selection dictated by the needs of 
the problem at hand.

For numerical calculations involving objects of
complicated shapes, it is convenient to follow
the practice of discretized integral-equation
solvers in computational 
engineering~\cite{Harrington1996},
by using 
\textit{localized} basis functions describing
elemental current distributions confined within
geometric regions of specific shapes (triangles,
tetrahedra, cubes, etc.) Bodies of complex shapes
may be approximated to arbitrary desired accuracy
as unions of these shapes (see insets of Figures
\ref{TorpedoFigure} and \ref{PinwheelFigure})
and the accuracy with which the current distribution
is represented may be systematically improved, at
cost of greater computational expense, by
refining the sizes of the elements.
Common examples include 
\textbf{(a)} RWG basis functions for surface 
currents~\cite{RWG1982}, which are localized in
triangles on discretized surface meshes such as
the mesh of Figure \ref{PinwheelFigure};
\textbf{(b)} SWG basis functions for volume
currents~\cite{SWG1984}, which are localized over
tetrahedra in discretized volume meshes such as
the mesh of Figure \ref{TorpedoFigure};
and 
\textbf{(c)} piecewise-constant basis functions
for volume currents~\cite{Polimeridis2015A},
which may be defined on uniformly-spaced voxel 
grids to allow efficient FFT-based convolutions.
(RWG and SWG functions are named for their
inventors~\cite{RWG1982,SWG1984}.)


The numerical calculations reported in this paper
were obtained using localized basis functions
(RWG and SWG functions for SIE and VIE methods, 
respectively). All calculations were performed 
using {\sc scuff-em}~\cite{scuff-em}
or {\sc buff-em}~\cite{buff-em}, 
both of which are free, open-source software 
packages available for download online.


\begin{figure}
\begin{center}
\resizebox{0.5\textwidth}{!}{\includegraphics{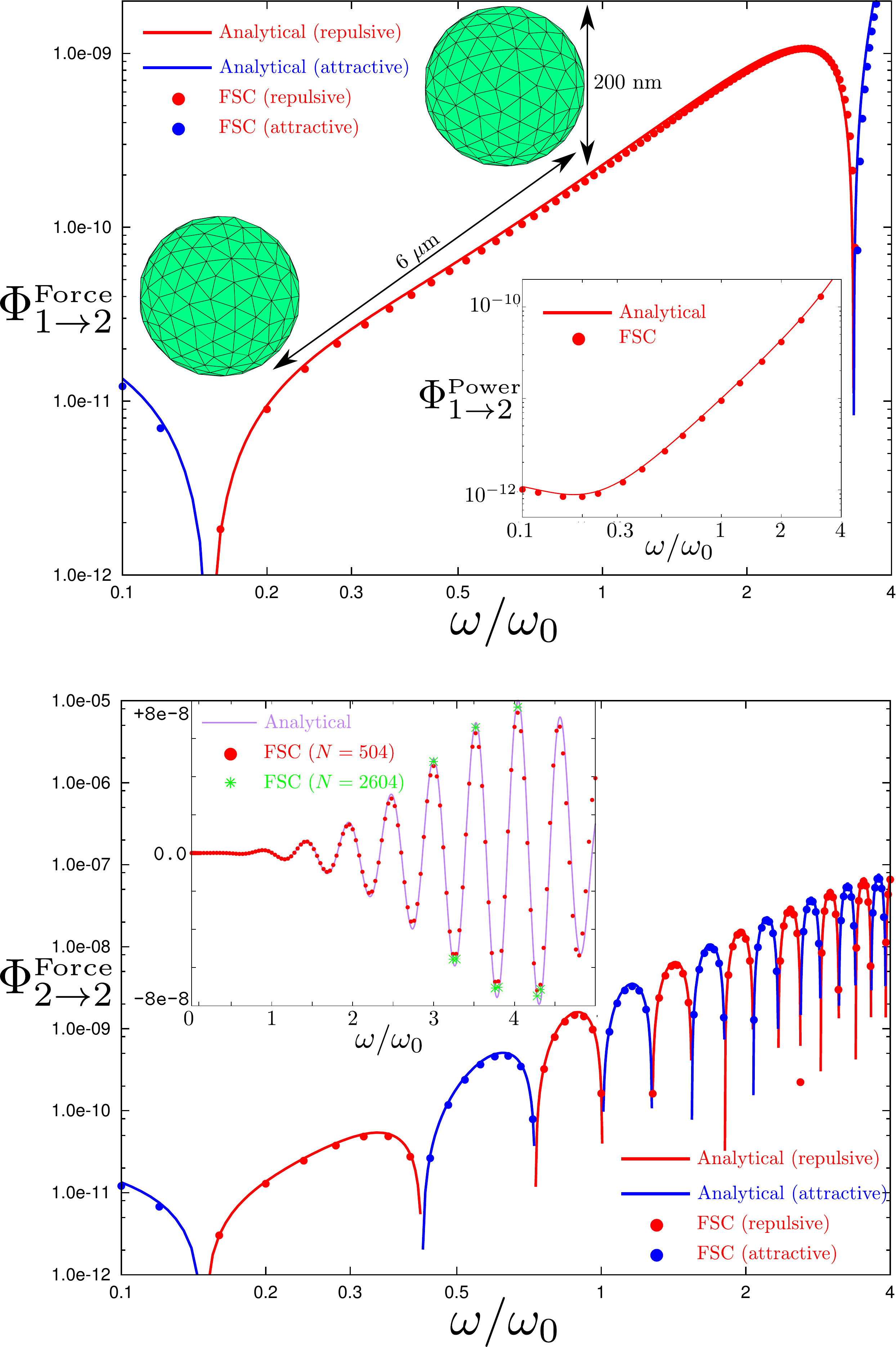}}
\end{center}
\caption{Frequency dependence of fluctuation-induced energy and momentum
flux between gold nanospheres of radius 100 nm separated by
a center-center distance of $6 \mu$m.
Solid lines indicate predictions of the analytical $T$-matrix formulas
of~\citeasnoun{Krueger2012}.
Filled circles denote FSC calculations for sphere surface meshes
with $N=504$ triangle edges (inset, not to scale).
Green stars in the inset of the lower plot denote FSC calculations
with finer surface meshes $(N=2604).$
\textit{Upper plot}: 
Flux (see text) of force on sphere 2 due to radiation from sphere 1.
\textit{Upper plot inset:} 
Flux of power into sphere 2 due to radiation from sphere 1.
\textit{Lower plot:} 
Flux of force on sphere 2 due to its own radiation,
graphed on log-log (main plot) and linear-linear (inset) scales.
Whereas the power flux is always positive and gently varying
with frequency, the force flux may be positive or negative
(attractive or repulsive) depending on the frequency,
with the self-force flux $\phi\sups{force}_{2\to 2}$ 
in particular exhibiting rapid oscillations.
The need to resolve these oscillations accurately
when integrating over frequency makes force calculations
significantly more costly than heat-transfer calculations.
\textit{Units:} Frequency is measured in
$\omega_0=3\cdot 10^{14}$ rad/sec.
Power fluxes $\phi\sups{Power}$ have units of watts / $\hbar \omega_0^2.$
Force fluxes $\phi\sups{Force}$ have units of nanoNewtons / $\hbar \omega_0^2.$
\red{Note that the quantities plotted here are temperature-independent
     \textit{fluxes}, which must be 
     weighted by the Bose-Einstein factor $\Theta(T,\omega)$
     and integrated over frequency
     to yield actual rates of energy and momentum transfer
     at a given temperature $T$}.
}
\label{GoldSpheresFigure}
\end{figure}
\section{Validation}
\label{ValidationSection}

Before using our new tools to predict new
fluctuation-induced interactions in previously-unexplored
geometries (Section \ref{ApplicationsSection}),
we first validate our methods by demonstrating that
they reproduce the results of~\citeasnoun{Krueger2012}
for the non-equilibrium Casimir force between homogeneous
spheres---among the few existing theoretical results
for momentum transfer between compact bodies out of
thermal equilibrium.

We consider homogeneous spheres of radius $R=100$~nm
separated by a distance of $d$=6 $\mu$m along
the $z$ axis (sphere $2$ lies above sphere $1$).
The spheres are composed of gold, \red{whose material
properties at the infrared frequencies relevant
for thermal radiation are adequately described by
a Drude-type dielectric function}~\cite{Krueger2012}:
\numeq{GoldPermittivity}
{
\epsilon\sups{gold}(\omega) =
   1-\frac{\omega_p^2}{\omega (\omega + i\gamma)}
}
with $\{\omega_p, \gamma\}=\{1.37\cdot 10^{16}, 5.32\cdot 10^{13}\}$
rad/sec.
The contribution of fluctuations in sphere $s$ to the
total $z$-directed force on sphere $d$ ($s,d\in\{1,2\}$) 
is given by
\numeq{FzIntegral}
{ F\subs{s$\to$d} =
   \int_0^\infty \Theta(T_s,\omega)
   \Phi\sups{Force}\subs{s$\to$d}(\omega) \, d\omega
}
where $T_s$ is the temperature of sphere $s$ (assumed constant
throughout the sphere) and the temperature-independent
force ``flux'' $\Phi\sups{Force}$ may be computed 
using analytical formulas given in~\citeasnoun{Krueger2012}
or numerically using FSC/FVC methods:
putting $Q=F_z$ in equation (\ref{CQC}g), we have
$\Phi\sups{Force}=\ExpVal{F_z}/\Theta(T_s,\omega)$.
The power transfer from sphere $s$ to sphere
$d$ is given by (\ref{FzIntegral}) with $\Phi\sups{Force}$
replaced by the power flux $\Phi\sups{Power}$.

Figure \ref{GoldSpheresFigure} plots
$\phi\sups{Force}_{1\to 2}$ (upper) and
$\phi\sups{Force}_{2\to 2}$ (lower) and
as computed using the numerical FSC solver
{\sc scuff-em}~\cite{scuff-em} (circles)
and using the analytical formulas of~\citeasnoun{Krueger2012}
(solid lines). For comparison,
we also plot $\phi\sups{Power}_{1\to 2}$
(inset of upper plot).
Red (blue) circles/lines denote
positive (negative) quantities, with positive force data
corresponding to repulsive forces.
The sphere surfaces are discretized into unions
of flat triangular panels (upper inset) with $N_\text{edge}=504$
total triangle edges per sphere; the total number of 
surface-current basis functions is $4N_\text{edge}$ (electric and 
magnetic currents on both spheres), so the 
dimension of the matrices in (\ref{CQC}g)
is $N\subt{BF}=2016$.

The agreement between numerical FSC data and the
analytical formulas is evidently excellent for all
quantities at all frequencies. In particular,
the FSC results accurately capture the rapid
oscillations of $\Phi\sups{Force}\subt{2$\to$2}(\omega)$
(main plot in lower panel),
which~\citeasnoun{Krueger2012} attributed to
constructive/destructive interference
between waves emitted by sphere 2 and
their reflections from sphere 1.
Closer scrutiny of the data on a linear-linear scale
(inset of lower panel) reveal slight discrepancies
between FSC and analytical data near peaks and troughs
of the oscillation; these are due to finite
meshing resolution and are reduced by repeating
FSC calculations with spheres meshed at
the finer resolution of $N_\text{edge}=2604$ interior
edges (green stars in lower inset), corresponding 
to matrices of dimension $N\subt{BF}=8256$ in (\ref{CQC}g).

Beyond validating the methods proposed in this paper,
Figure \ref{GoldSpheresFigure} lends intuition to
the discussion of Section \ref{WhyHardSection} regarding
the greater computational cost of force calculations
as compared to power calculations. The force
integrand in (\ref{FzIntegral}) exhibits sign changes
for both the $1\to 2$ and $2\to 2$ cases---with
particularly violent oscillations in the latter
case---requiring large numbers of integrand samples,
each involving costly evaluation of the matrix-trace
formulas (\ref{CQC}g), to yield accurate numerical
estimates of the total integrated force.
In contrast, the power integrand (inset of upper plot)
is always positive and varies slowly with frequency,
allowing the frequency integral to be evaluated
at modest computational cost.

\section{Applications}
\label{ApplicationsSection}

\subsection{The photon torpedo: Self-propulsion of warm 
            inhomogeneous nanoparticles in a cold environment}
\label{PhotonTorpedoSection}

\begin{figure}
\begin{center}
\resizebox{0.5\textwidth}{!}{\includegraphics{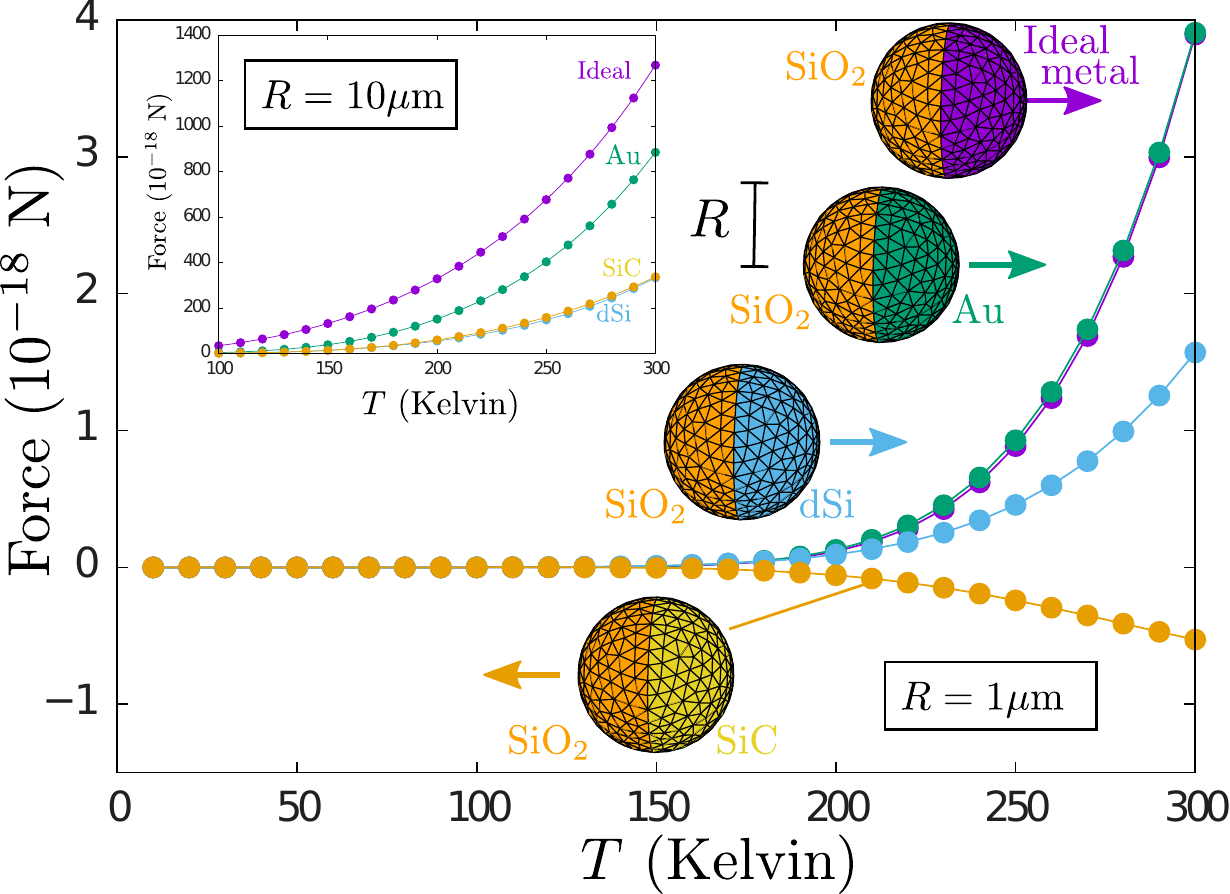}}
\end{center}
\caption{Thermal self-propulsion of warm (temperature $T$)
``photon torpedoes'' in a cold (0 K) environment.
The torpedoes are ``Janus'' particles~\cite{Walther2013}
consisting of SiO$_2$ spheres of radius $R$ with
one full hemisphere replaced by an impostor
substance: ideal plasmonic metal (purple),
real gold (green), doped silicon (cyan), or silicon
carbide (yellow).
Simple arguments (see text) suggest
that the self-propulsion force experienced by the 
particle should be directed toward the impostor
hemisphere; for particles of radius $R=1\,\mu$m
(main figure), this prediction is borne out for 
the three metallic impostor substances, but
the SiO$_2$-SiC particle self-propels in the opposite
direction (lowermost curve in main plot).
This phenomenon reverses itself for larger particles
with $R=10\,\mu$m (inset); now the force
is directed toward the impostor region for
all material combinations. The magnitude
of the force is on the order of $10^{-18}$ N
for the $1 \mu$m particles and scales roughly
with the volume of the particle, increasing 
approximately 1,000 fold for $10\,\mu$m-radius
particles.
}
\label{TorpedoFigure}
\end{figure}

Figure \ref{TorpedoFigure} plots, as a function of temperature,
the self-propulsion force on various hybrid particles
designed to realize the notion of a ``photon torpedo.''
The idea is to surround a mass of homogeneous, thermally
radiating material---\red{silicon dioxide in this case}---with
a \textit{partial} reflective coating;
for example, one might consider replacing the lower hemisphere of
a 1 $\mu$m-radius SiO$_2$ sphere with a metallic substance
(inset).
Then thermally-emitted photons
radiated in the direction of the lower hemisphere are
reflected by the metallic region, ensuring a net 
upward-directed stream of momentum carried away by the 
thermally-emitted photons; to conserve momentum, the particle
must \textit{recoil} in the downward direction---that is,
the direction of the reflective coating. Micron-scale
``Janus'' particles of this sort may be readily
fabricated via standard techniques and are a focus
of current interest in nanophotonics 
research~\cite{Walther2013, Ilic2017};
here we explore their non-equilibrium self-propulsion
properties.

\begin{figure}
\begin{center}
\resizebox{0.5\textwidth}{!}{\includegraphics{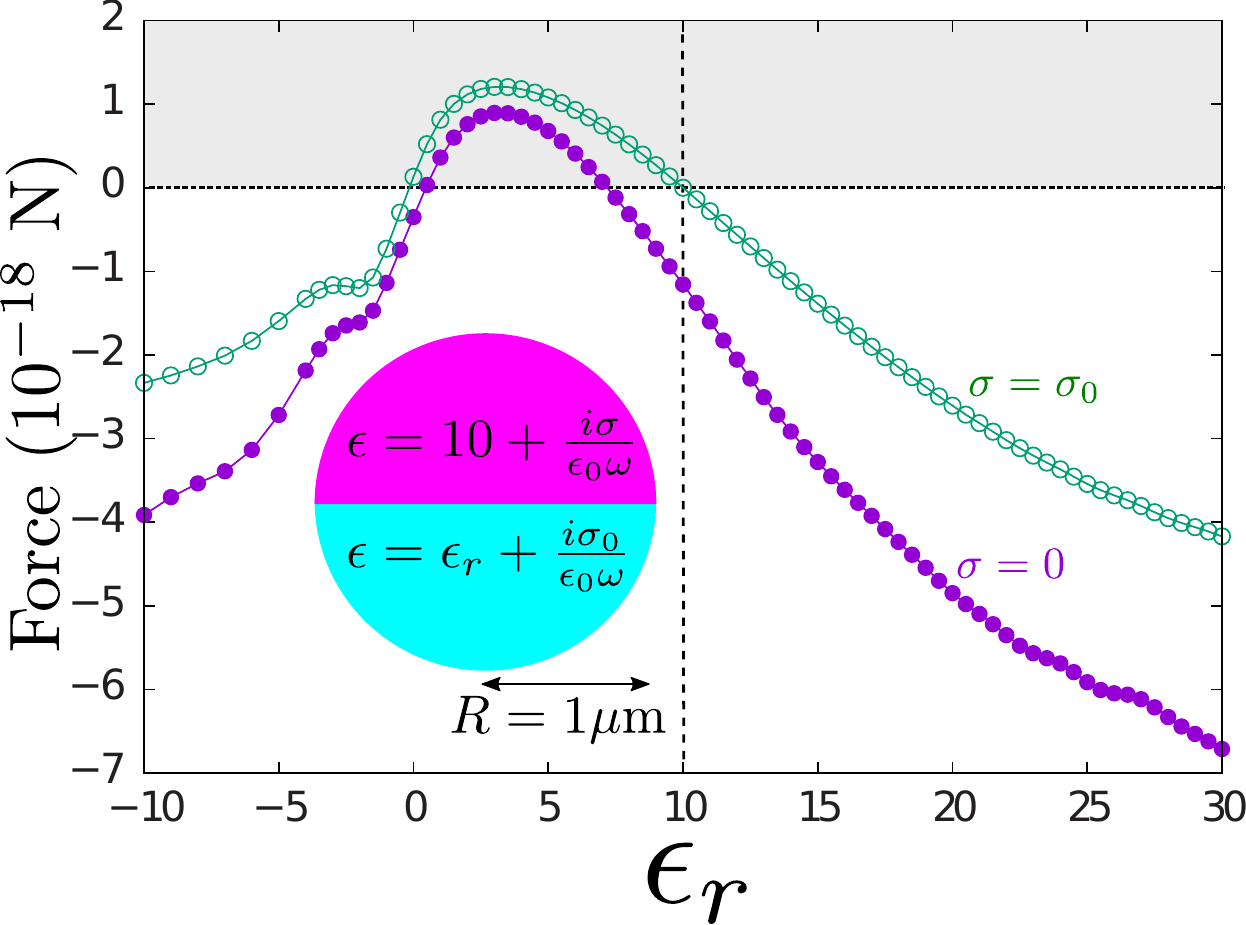}}
\caption{Self-propulsion vs. dielectric contrast of warm (300 K)
Janus particle in 0K background. The conductivity of the lower 
hemisphere is fixed at $\sigma_0=(Z_0\cdot 1\,\mu\text{m})$,
while the real part of its relative dielectric constant varies
over the range $\epsilon_r\in[-30,10]$.
The upper hemisphere has real relative dielectric constant 
fixed at 10 and either vanishing conductivity (filled purple 
circles) or conductivity $\sigma_0$ (hollow green
circles). By symmetry the force on the particle must vanish when the
upper and lower media are identical (intersection of dashed lines),
but for other media combinations the magnitude and sign
of the force vary in non-intuitive ways.
}
\label{JanusMapFigure}
\end{center}
\end{figure}
In Figure \ref{TorpedoFigure} we have investigated this
concept for Janus particles of radius $R=\{1,10\}\mu$m
consisting of an SiO$_2$ hemisphere paired
with hemispheres of various materials:
\textbf{(a)} an idealized lossless metal (purple)
described by the plasmonic dielectric function
$$ \epsilon\sups{ideal metal}(\omega) = 1-\pf{\omega_p}{\omega}^2$$
with $\omega_p=3\cdot 10^{16}$ rad/sec;
\textbf{(b)} real gold (Au) (green) with the dielectric function
of Eq. (\ref{GoldPermittivity}),
\textbf{(c)} doped silicon (dSi) (cyan), 
and
\textbf{(d)} silicon carbide (SiC) (yellow).
The figure plots the magnitude and sign of the self-propulsion
force experienced by the Janus particle versus its temperature $T$
in a background environment maintained at $T=0$ K.
\begin{figure}
\begin{center}
\resizebox{0.5\textwidth}{!}{\includegraphics{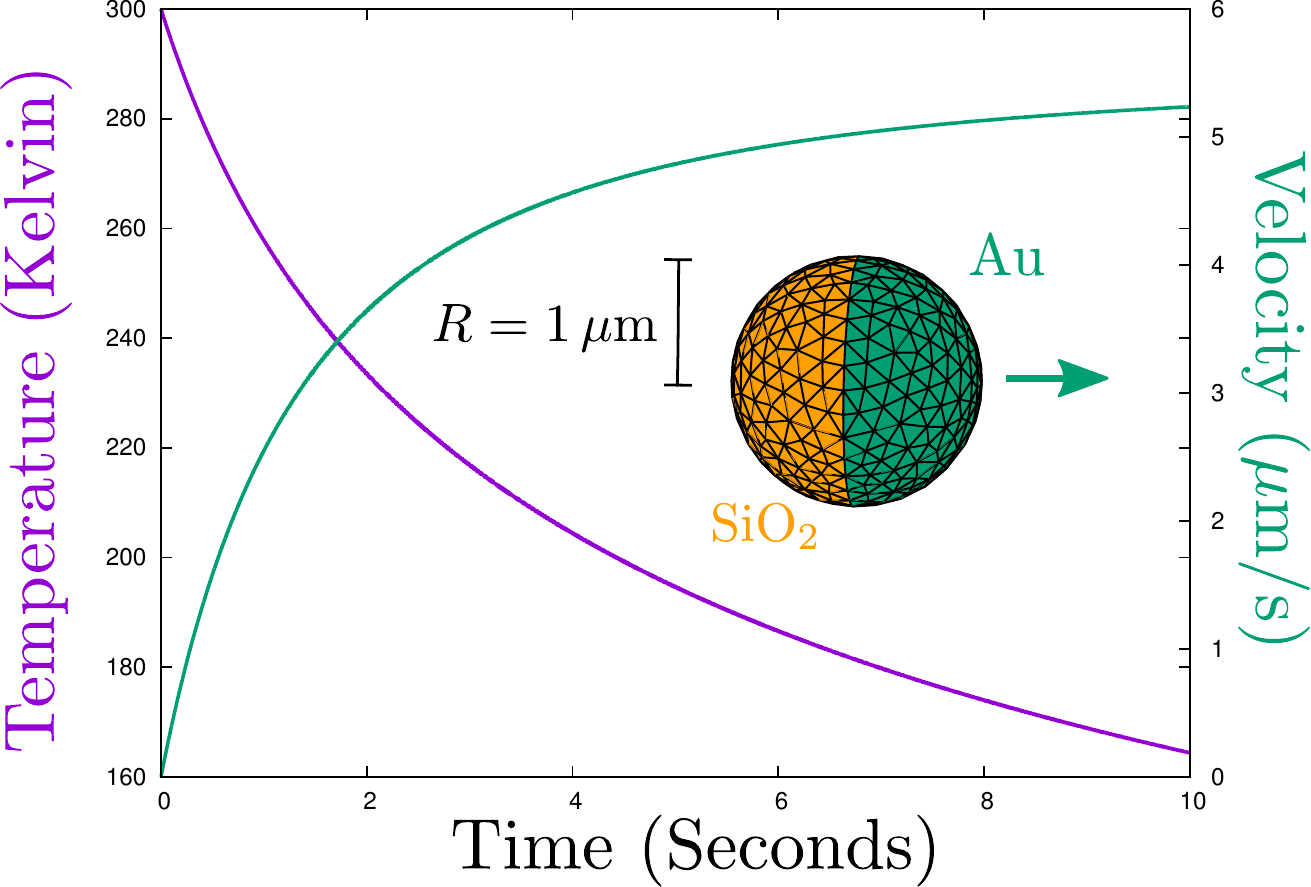}}
\end{center}
\caption{Temporal evolution of the temperature (purple) and velocity
(green) on a 1-$\mu$m gold-SiO$_2$ Janus-particle photon torpedo
heated to a temperature of 300 K and released into a cold (0K) 
environment. The torpedo ramps up to a terminal linear velocity
of $\sim 5$ $\mu$m /s over an interval of $\sim 10$ s.}
\label{PTTerminalVelocityFigure}
\end{figure}
For the three metallic substances, the particle
experiences a self-propulsion force
directed toward the metallic hemisphere
in keeping with the intuitive prediction of the photon-torpedo
picture; this is true both for particles of radius $R=1\,\mu$m (main
figure) and for larger particles with $R=10\,\mu$m (inset).
On the other hand, the behavior is more subtle when the
metallic material is replaced by the dielectric
insulator SiC: \red{now the force is directed
toward the SiO$_2$ hemisphere (that is, opposite the direction
expected for a ``photon torpedo'' and observed for the metallic
hemispheres) for $R=1\,\mu$m particles, but toward the 
SiC hemisphere (i.e. in the ``torpedo-like'' direction)
for $R=10\,\mu$m.}

The difficulty of predicting the sign of
thermal-propulsion forces
is further illustrated by Figure \ref{JanusMapFigure},
in which we have plotted the self-propulsion force
of a 300K Janus particle in a 0K background
as a function of the dielectric contrast between
upper and lower hemispheres. The lower hemisphere
has relative permittivity
$\epsilon=\epsilon_r +\frac{i\sigma_0}{\epsilon_0\omega}$
with conductivity fixed at $\sigma_0=(Z_0\cdot 1\mu$m$)^{-1}$
and varying real part $\epsilon_r$.
The upper hemisphere has fixed real relative
dielectric constant $10$ and conductivity 
$\sigma=0$ (purple) or $\sigma_0$ (green).
The self-propulsion force vanishes by symmetry
when the upper and lower hemispheres are identical 
(intersection of dashed lines); for other values of 
the dielectric 
constrast it is difficult to predict the magnitude
and even the \textit{sign} of the force.

We conclude that the design of 
self-propelling nanoparticles is subtle; heuristic
design intuition may prevail in some cases, but
\red{reliable predictions require rigorous
numerical tools like those presented in this paper.}

\textit{Terminal velocities.}
A particle of temperature $T\supt{P}$ released into an
environment of temperature $T\sups{env}$ will eventually
equilibrate through (among other mechanisms) the emission 
or absorption of thermal radiation. Neglecting conductive 
and convective heat transfer, the temporal evolution of the 
particle temperature is given by
\begin{subequations}
\begin{align}
 \frac{dT}{dt} &= - \frac{1}{\kappa} H(T)
\label{dTdt}
\intertext{with $\kappa$ the heat capacity and 
           $H=\frac{dU}{dt}$ the temperature-dependent rate
           of energy emission/absorption due to thermal radiation.
           Solving (\ref{dTdt}) with boundary condition
           $T(t=0)=T\supt{P}$ yields a function $T(t)$
           describing the time evolution of the particle
           temperature; at the equilibration time $t^*$
           defined by $T(t^*)=T\sups{env}$ the particle has
           exhausted its ``fuel'' and thereafter exhibits no 
           net radiative exchange of energy or momentum with
           the medium. Neglecting drag forces and other
           dissipative effects, the particle then remains 
           in motion at fixed linear and angular velocities 
           $\{\vb v, \vbOmega\}\sups{terminal}$ given by
          }
 \vb v\sups{terminal}    &= \frac{1}{m}\int_0^{t^*} \vb F\Big(T(t)\Big) \, dt \\
 \vbOmega\sups{terminal} &= \frac{1}{I}\int_0^{t^*} \bmc T\Big(T(t)\Big) \, dt
\end{align}
\label{TerminalVelocities}%
\end{subequations}
with $m, I$ the particle mass and moment of inertia.
The temperature-dependent rates of energy and momentum exchange
$\{H, \vb F, \mc T\}(T)$ may be computed for arbitrary particles
using the methods discussed in Section II, and we may use
this information to determine the temperataure trajectory $T(t)$
and compute the terminal velocities of warm nanoparticles 
released into vacuum.
Figure \ref{PTTerminalVelocityFigure} shows the time-varying
temperature (purple curve) and velocity (green curve) of the
gold-SiO$_2$ Janus-particle photon torpedo of
Figure \ref{TorpedoFigure}, assuming a heat capacity of
$\kappa$=800 J/(kg K)~\cite{Krueger2012}
an average mass density of $\rho\approx$ 3 g/cm$^3$,
and an environment temperature $T\sups{env}=0$ K.
The torpedo ramps up to a terminal linear velocity
of $\sim 5$ $\mu$m/s over an interval of $\sim 10$ s.

\textit{Dynamics on microscopic length scales but macroscopic time scales.}
A curious feature of the self-propulsion dynamics of warm micron-size
bodies is the coexistence of microscopic length scales with macroscopic 
time scales. Indeed, forces on the order of $10^{-19}$ Newtons acting 
on particles of mass on the order of $10^{-14}$ kg yield accelerations 
on the order of 10 $\mu$m/s$^2$. An experimentalist observing the
resulting motion through a microscope could view or video-record
the trajectories in real time.

\textit{Order-of-magnitude comparisons of force and acceleration
mechanisms.}
For micron-scale particles at 300 K in 0K backgrounds,
the self-propulsion forces of
Figures \ref{TorpedoFigure}-\ref{PTTerminalVelocityFigure}
are on the order of $\sim 10^{-20}$ N, yielding
accelerations on the order of $\sim 1\,\mu$m/s.
By comparison, typical forces felt by micron-scale 
particles in laser optical traps are on the order
of $\sim 10^{-12}-10^{-10}$ N~\cite{Hansen1999,MaiaNeto2000},
while the gravitational acceleration at the
Earth's surface is $\sim 10$ m/s.

\begin{figure}
\begin{center}
\resizebox{0.5\textwidth}{!}{\includegraphics{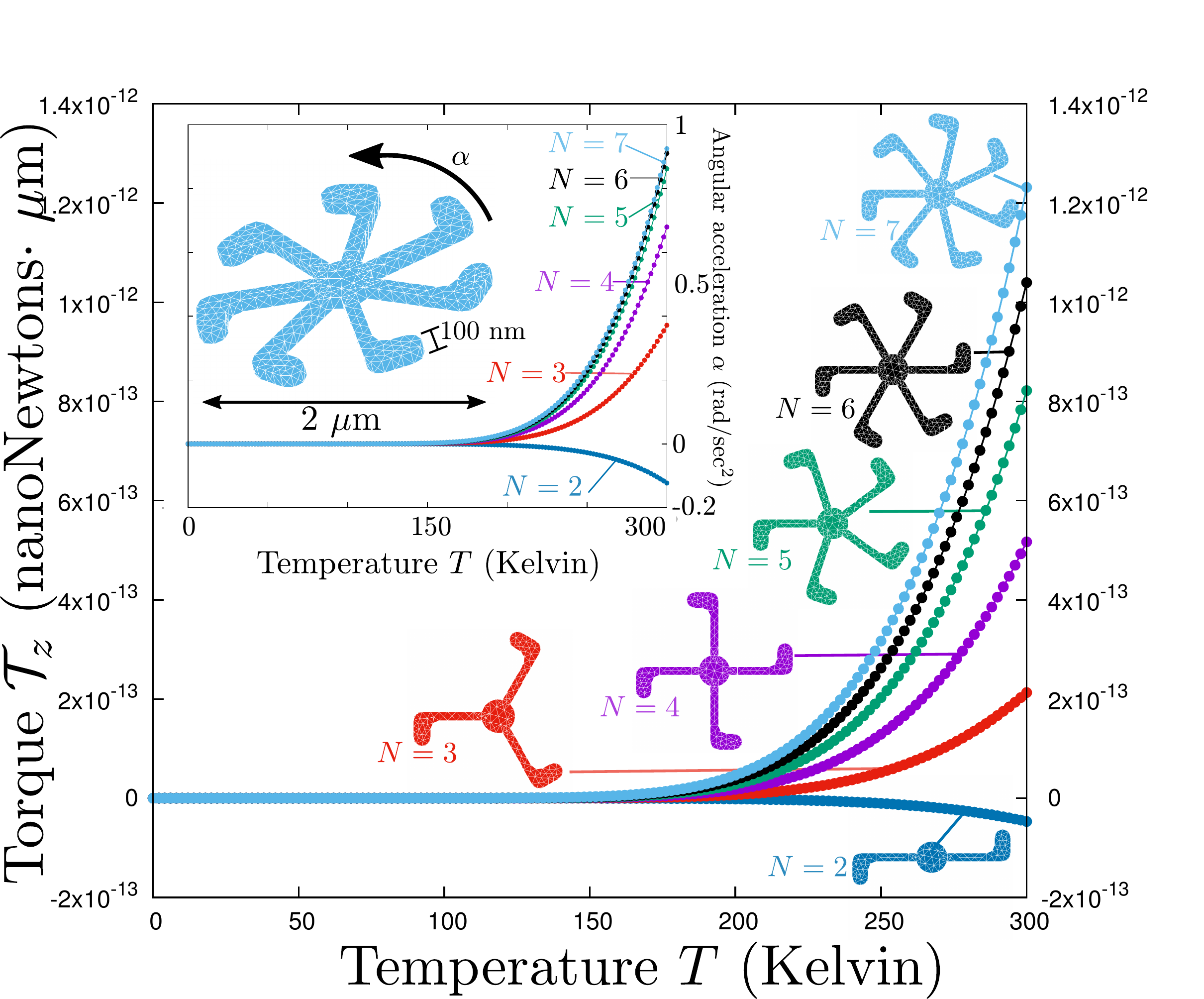}}
\end{center}
\caption{Thermal self-rotation of Rytov pinwheels in a cold
         environment. \textit{Main figure:} Self-rotation
         torque $\mc T_z$ vs. pinwheel temperature $T$.
         The magnitude of the torque at $T$=300 K
         grows steadily with the number of pinwheel arms $N$,
         increasing by roughly a factor of 4 as $N$ varies from
         2 to 7.
         \textit{Inset:} Angular acceleration $\alpha=\mc T_z/I_z$
         with $I_z$ the moment of inertia about the $z$ axis. Now
         all curves approximately collapse onto a single universal
         trajectory, yielding a macroscopic angular acceleration
         of $\sim$ 0.05 rad/sec$^2$ at 300 K.\red{Increased font size
         of inset axis labels.}
}
\label{PinwheelFigure}
\end{figure}

\subsection{The Rytov pinwheel: Self-rotation of warm chiral 
            nanoparticles in a cold environment}
\label{RytovPinwheelSection}

Figure \ref{PinwheelFigure} plots the self-rotation torque $\mc T_z$
experienced by chiral nanoparticles---``Rytov pinwheels''---of
various temperatures $T$ in a cold (0K) environment.
The pinwheels consist of gold sheets of thickness 100 nm
etched into the shape of a central disc from which
emanate $N=\{2,3,4,5,6,7\}$ arms extending to an outer radius
of \red{$1\,\mu$m}---an object that could be fabricated by
standard lithographic and etching techniques. The net loss
of angular momentum
due to the asymmetric polarization of thermal radiation
imparts to the pinwheel a recoil torque whose magnitude
grows steadily with $N$ (main figure), increasing nearly
sixfold as $N$ varies over the range $[3,7]$. Interestingly,
the $N=2$ pinwheel exhibits self-rotation
in the direction opposite that of all other pinwheels.

\begin{figure}
\begin{center}
\resizebox{0.5\textwidth}{!}{\includegraphics{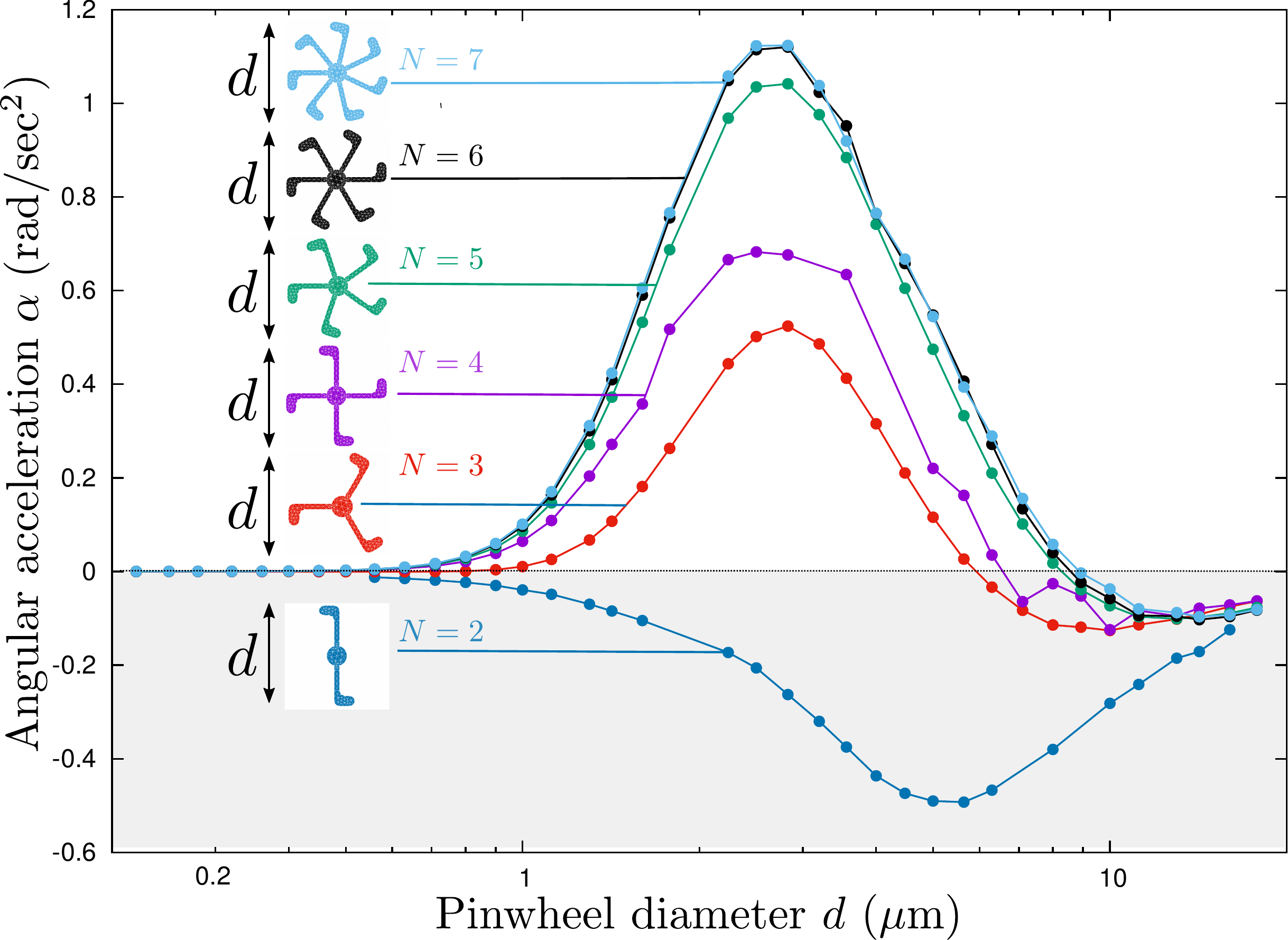}}
\end{center}
\caption{Angular acceleration $\alpha=\mc T_z/I_z$
         vs. diameter $d$ for warm ($T=300 $K)
         gold pinwheels with various numbers of arms $N$
         in a cold ($T=0$ K) environment.
         For all $N$ we find an optimal diameter
         $d\sups{opt}\in [2,6] \mu$m at which the acceleration
         is maximized. For $N=2$ the acceleration
         is negative (pinwheel spins clockwise as seen
         looking into the page) for all diameters $d$.
         For $3\le N \le 7$ the 
         acceleration is positive (pinwheel spins counter-clockwise)
         for diameters near $d\sups{opt}$ but reverses sign for
         larger $d$.
        }
\label{PinwheelVsDFigure}
\end{figure}

Dividing by the moment of inertia
$I_z=\int_V (x^2+y^2)\rho\,dV$ to
yield values of the angular acceleration $\alpha=\mc T_z/I_z$
collapses the $N=\{5,6,7\}$ curves onto a single
trajectory, with the $N=4$ case also falling within 
$\sim$ 20\% (inset). This suggests the existence of a
limiting angular acceleration attained by self-rotating
chiral particles in the $\mc I_z\to\infty$ limit.

How does this limiting self-acceleration vary with the
absolute size of a particle of fixed shape and aspect ratio?
Figure \ref{PinwheelVsDFigure} plots angular acceleration
vs. pinwheel diameter for pinwheels at fixed temperature
$T=300$ K in a 0 K environment; the diameter-$d$ pinwheel 
with $N$ arms is simply the $N$-arm pinwheel of
Figure \ref{PinwheelFigure} with all linear dimensions
scaled uniformly by $d/2\,\mu$m so that aspect ratios
are preserved (in particular, the ratio of thickness
to diameter remains fixed at $1/20$, as in 
the upper inset of Figure \ref{PinwheelFigure}.)
For all values of $N$ we find a well-defined 
optimum particle diameter in the range
$d\sups{opt}\in [2,6] \mu$m at which self-acceleration is
maximized. (The value of $d\sups{opt}$ shifts
to larger values as the pinwheel temperature decreases.)

As in Figure \ref{PinwheelFigure}, for most values of $d$
the $N=2$ pinwheel self-rotates in the direction opposite
that of the $N>2$ pinwheels; this is true in particular
for particles at or near the optimal diameter $d\sups{opt}$.
On the other hand, as $d$ increases beyond $d\sups{opt}$
the self-acceleration of the $N=3$ and higher pinwheels 
decreases in magnitude and eventually changes sign,
so that for $d\gtrsim 8\,\mu$m all pinwheels 
join the $N=2$ case in rotating counter-clockwise.
Partial insight into this curious phenomenon is afforded
by considering the (deterministic) angular-momentum
absorption profile of the pinwheels under irradiation
by a circularly-polarized beam 
(see Figure \label{GearPinwheelMechanismFigure} and
discussion below).

Like the photon torpedo, the self-rotation dynamics of the
Rytov pinwheel merge microscopic length scales with macroscopic
time scales: these micron-scale bodies spin at human-scale
angular frequencies on the order of $\sim$ 1 rad/sec, motions
readily observable under a microscope in real time.
\subsection{The non-contact thermal microgear:
            Temperature-dependent sign of angular-momentum transfer
            between warm and cold bodies}
\label{MicroGearSection}

\begin{figure}
\begin{center}
\resizebox{0.5\textwidth}{!}{\includegraphics{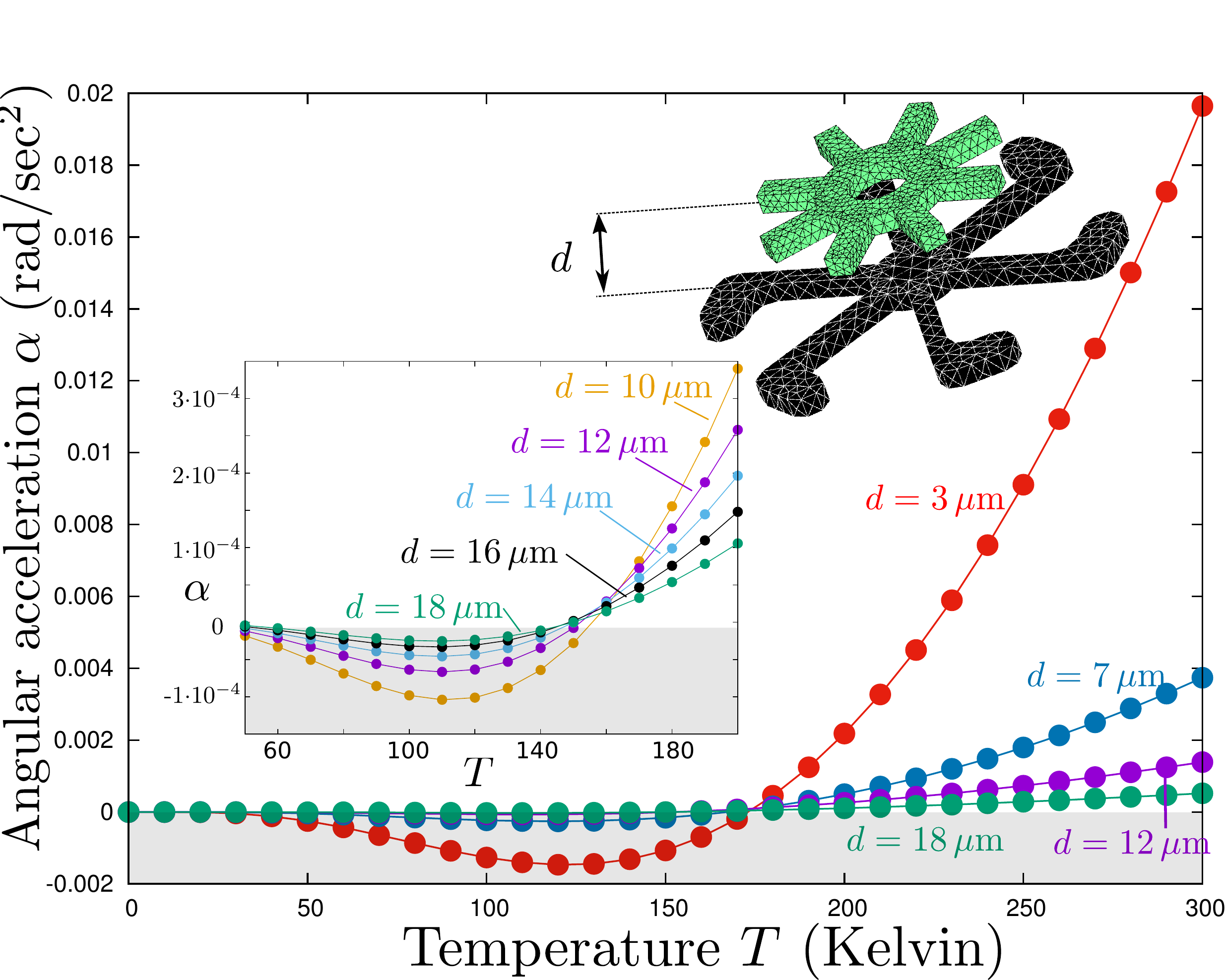}}
\end{center}
\caption{Torque on a cold (0 K) gold microgear induced by
         absorption of angular momentum from thermal radiation
         emitted by the $N=6$ Rytov pinwheel of
         Figure \ref{PinwheelFigure} at various pinwheel temperatures
         $T$ in a cold (0 K) environment. The torque exhibits
         an unexpected change in sign at a pinwheel temperature 
         near $T$=160 K. Although the magnitude of the
         torque is greatest at the smallest gear-pinwheel
         separation distance of $d$=3 $\mu$m, the sign-reversal
         effect persists even at much greater distances (inset).
         \red{Increased font size of axis labels.}
}
\label{GearPinwheelFigure}
\end{figure}
Can the angular momentum radiated by a warm chiral body
be captured by a nearby cold body to induce rotation?
Figure \ref{GearPinwheelFigure} plots the angular acceleration
of a cold (0 K) gold microgear, positioned at various
(center-center) distances $d$ above the $N=6$ pinwheel 
of Figure \ref{PinwheelFigure}, as a function of the 
pinwheel temperature $T$ in a
cold (0 K) environment. 
Like the pinwheel, the microgear is etched from a $1$-$\mu$m
thick sheet of gold; its outer radius is approximately
$6$ $\mu$m.
The torque exhibits a surprising non-monotonic
temperature dependence, reversing
sign at a $d$-independent temperature near $T=$160 K. 
Although the magnitude of the torque is largest at small
gear-pinwheel separation distance $d=3$ $\mu$m, the 
sign-reversal effect persists at all separation distances 
considered, including separations as large as $d=18$ $\mu$m
(inset).
\begin{figure}
\begin{center}
\resizebox{0.5\textwidth}{!}{\includegraphics{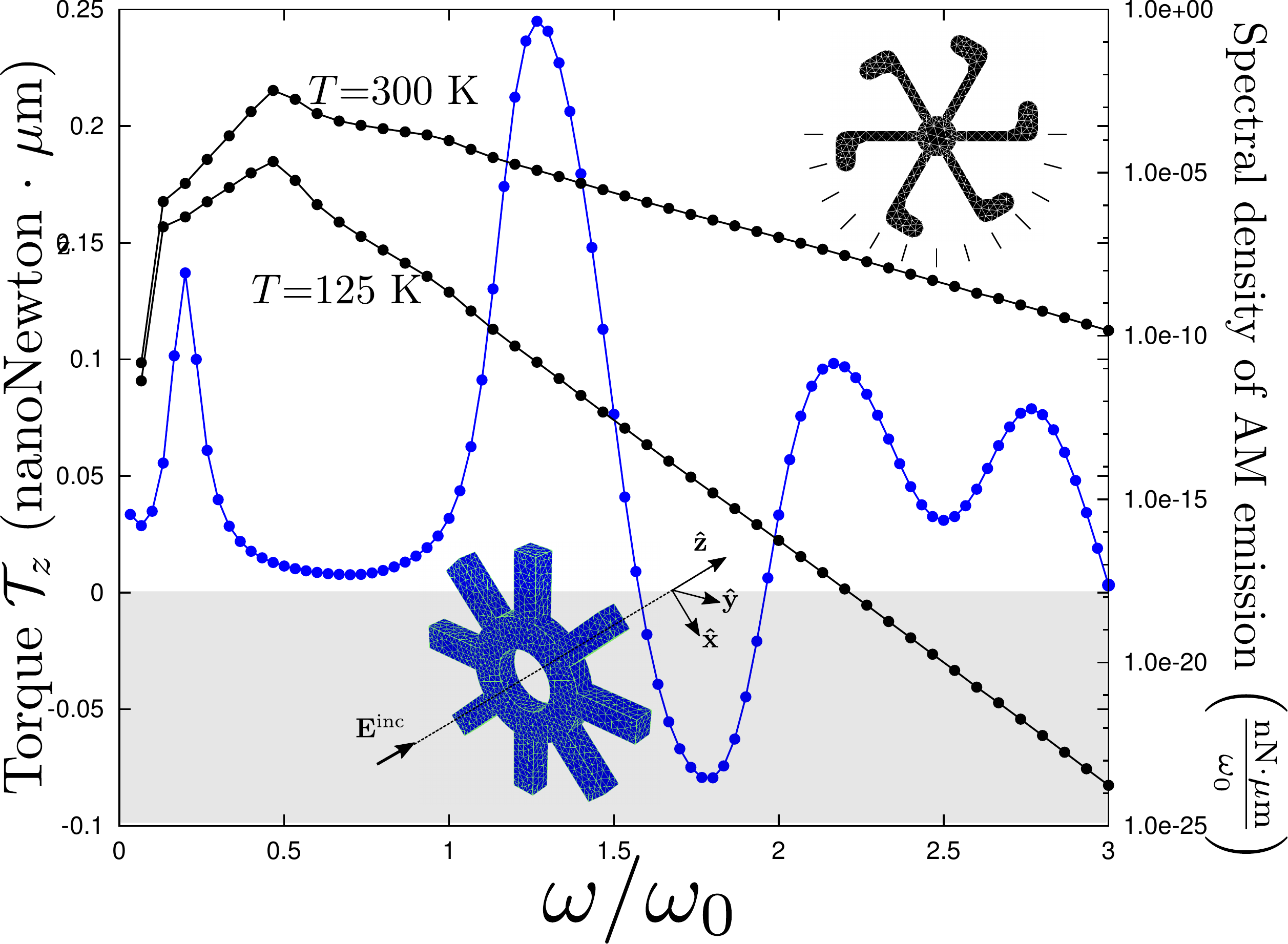}}
\end{center}
\caption{Elucidating the mechanism of the temperature-dependent
         sign of the torque in Figure \ref{GearPinwheelFigure}.
         environment. The purple curve shows the rate of angular
         momentum (AM) absorption (radiation torque) for the isolated
         microgear illuminated by a circularly-polarized plane
         wave (lower inset) vs. the plane-wave angular frequency 
         $\omega$, measured in units of $\omega_0=3\cdot 10^{14}$ rad/sec.
         There exists a frequency window in the range
         $1.5\omega_0<\omega<2\omega_0$ within which the
         gear extracts angular momentum of the opposite sign
         from the polarization of the bare plane wave.
         The blue and green curves show the spectral density of
         angular momentum carried away by thermal radiation
         from an isolated $N=6$ Rytov pinwheel (upper inset)
         at pinwheel temperatures $T$=125 K (green) and $T=$300 K (blue).
         Depending on the temperature of the pinwheel, the
         gear's window of opposite-sign angular-momentum absorption
         is or is not substantially excited, yielding the strongly 
         temperature-dependent torque of Figure \ref{GearPinwheelFigure}.
        }
\label{GearPinwheelMechanismFigure}
\end{figure}

The origins of the temperature-dependent sign of this
\textit{fluctuation-induced} torque may be understood
by considering the angular-momentum absorption profile
of the gear in a \textit{deterministic} setting.
Figure \ref{GearPinwheelMechanismFigure} plots
the torque on the isolated gear irradiated by
a circularly-polarized plane wave
$\vb E\sups{inc}(\vb x)=
 \frac{E_0}{\sqrt{2}}\left(\vbhat{x} + i\vbhat{y}\right)e^{i\omega z/c}$
with $E_0=1$ V / $\mu$m.
In this figure, the purple curve shows the rate of angular
momentum (AM) absorption (radiation torque) for the isolated
microgear illuminated by a circularly-polarized plane
wave (lower inset) vs. the plane-wave angular frequency 
$\omega$, measured in units of $\omega_0=3\cdot 10^{14}$ rad/sec.
There exists a frequency window in the range
$1.5\omega_0<\omega<2\omega_0$ within which the
gear extracts angular momentum of the opposite sign
from the polarization of the bare plane wave.
The blue and green curves show the spectral density of
angular momentum carried away by thermal radiation
from an isolated $N=6$ Rytov pinwheel (upper inset)
at pinwheel temperatures $T$=125 K (green) and $T=$300 K (blue).
[This is just the negative of the integrand in
(\ref{CQC}g) for the pinwheel self-torque.]
Depending on the temperature of the pinwheel, the
gear's window of opposite-sign angular-momentum absorption
is or is not substantially excited, yielding the strongly 
temperature-dependent torque of Figure \ref{GearPinwheelFigure}.

\section{Conclusions} 
\label{ConclusionsSection}

The subtle, intuition-confounding behavior on display
in Figures \ref{TorpedoFigure}, \ref{PinwheelVsDFigure},
and \ref{GearPinwheelFigure} testifies to the need for
rigorous quantitative approaches to the modeling of
non-equilibrium fluctuation-induced interactions. In
contrast to equilibrium Casimir phenomena---for which
simple pictures such as the proximity-force
approximation (PFA)~\cite{Derjaguin1960} suffice
in many cases (though certainly not all~\cite{Gies1996})
to lend qualitatively correct insight,
and simple approximations such as the leading terms in 
scattering-theoretic series expansions often suffice to
capture key qualitative features~\cite{Emig2007}---the
intricate nature of non-equilibrium phenomena often
invalidates na\"ive intuitive approaches and demands
the full power of the unique new computational apparatus
we have supplied. Indeed, it is difficult to imagine
how the answers even to binary questions such as the
direction of self-propulsion for the photon torpedoes
of Figure (\ref{TorpedoFigure}), or the sign of the
torque for the Rytov pinwheels of Figures (\ref{PinwheelFigure})
and (\ref{PinwheelVsDFigure}), could have been
predicted with \textit{any} confidence
based on \textit{a priori} intuition or
back-of-envelope estimates---nor addressed with
any existing paradigm for predicting non-equilibrim
interactions.
Of course, perhaps with time the various novel
phenomena predicted by our numerical tools
will lend themselves to new sorts of intuitive
pictures, and new PFA-like approaches to 
nonequilibrium forces and torques will arise;
we hope the efficient new algorithms we have
presented here, together with our free, open-source
software implementations~\cite{scuff-em,buff-em},
will be of service in speeding the arrival of
this new understanding.

It is interesting to note that the characteristic motion
of micron-scale self-propelling and self-rotating particles
involves microscopic length scales but macroscopic time scales;
typical linear accelerations and terminal velocities
are on the order of microns/second$^2$ and microns/second,
while typical terminal angular velocities are on the order of
radians per second.
An experimentalist observing or filming these particles
through a microscope would be able to follow their trajectories
in real time.
%

In a different vein, we noted in Section \ref{WhyHardSection}
that one of the problems tackled in this paper---the self-force
on a macroscopic, continuum dielectric body resulting from
its own radiation---has a microscopic analogue in the
well-known problem of the radiative reaction force felt by
an accelerating \textit{pointlike}
particle~\cite{Ford1991, Yaghjian2008, Griffiths2010}.
Theoretical approaches to this problem~\cite{Moniz1977, Jackson1999}
often proceed by effectively discretizing a finite-volume model
of an accelerating particle and writing equations similar
to our equation (\ref{ForceIntegral}) to estimate 
radiative reaction force. However, to our knowledge all
previous studies of this sort have been \textit{analytical}
efforts based on hand calculation, for which restrictive 
approximations such as rigidity and spherical symmetry---limiting
consideration to nonrelativistic motion---are mandatory to render
progress tractable. It may be of interest to apply a version of
the formalism we have developed here---perhaps with localized 
basis functions and meshed geometries capable of faithfully
capturing particle shape distortions in relativistic 
motion---to rigorous numerical studies of radiation reaction
beyond the non-relativistic regime.



%

\end{document}